\documentclass[prl,twocolumn,superscriptaddress]{revtex4-1}

\usepackage{graphicx}
\usepackage[colorlinks=true,linkcolor=blue]{hyperref}
\usepackage{amsmath} 
\usepackage[usenames]{color}
\usepackage{dcolumn}

\begin{document}

\newcommand{\Ventry}{\ensuremath{V_{\mathrm{entry}}}}
\newcommand{\VentryDC}{\ensuremath{\Ventry^{\mathrm{DC}}}}
\newcommand{\VLE}{\ensuremath{\Ventry^{\mathrm{load}}}}
\newcommand{\Vexit}{\ensuremath{V_{\mathrm{exit}}}}
\newcommand{\Ef}{\ensuremath{E_{\mathrm{F}}}}
\newcommand{\tload}{\ensuremath{T_{\mathrm{load}}}}

\title{Quantum dot  state initialization by control of tunneling rates}

\author{Tobias Wenz}
\affiliation{Physikalisch-Technische Bundesanstalt (PTB), Bundesallee 100, 38116 Braunschweig, Germany}
\author{Jevgeny Klochan}
\affiliation{Department of Physics, University of Latvia, Jelgavas street 3, LV 1004 Riga, Latvia}
\author{Frank Hohls}
\email{frank.hohls@ptb.de}
\affiliation{Physikalisch-Technische Bundesanstalt (PTB), Bundesallee 100, 38116 Braunschweig, Germany}
\author{Thomas Gerster}
\affiliation{Physikalisch-Technische Bundesanstalt (PTB), Bundesallee 100, 38116 Braunschweig, Germany}
\author{Vyacheslavs Kashcheyevs}
\email{slava@latnet.lv}
\affiliation{Department of Physics, University of Latvia, Jelgavas street 3, LV 1004 Riga, Latvia}
\author{Hans W. Schumacher}
\affiliation{Physikalisch-Technische Bundesanstalt (PTB), Bundesallee 100, 38116 Braunschweig, Germany}

\begin{abstract}
We study the loading of electrons into a quantum dot with dynamically controlled tunnel barriers. We introduce a method to measure tunneling rates for individual discrete states and to identify their relaxation paths. 
Exponential selectivity of the tunnel coupling enables loading into specific quantum dot states by tuning independently energy and rates.
While for the single-electron case orbital relaxation leads to fast transition into the ground state, for electron pairs triplet-to-singlet relaxation is suppressed by long spin-flip times. This enables the fast gate-controlled initialization of either a singlet or a triplet electron pair state in a quantum dot with broad potential applications in quantum technologies. 
\end{abstract}

\maketitle 

The ability to initialize a discrete quantum state by coupling to a cold external environment is an essential resource for many quantum technologies, providing, for example, a template to encode quantum information, an initial pure state for quantum sensing, or a supply of ancillary qubits for quantum error correction \cite{DiVincenzo2000,Knill2005}. 
In solid-state quantum platforms such as semiconductor quantum-dots (QDs),
initialisation strategies vary from simple cooling by spontaneous relaxation \cite{Loss1998,DiVincenzo2000} to control of tunnelling to adjacent electron reservoir(s) via energy alignment \cite{Petta2005} or Pauli blockade~\cite{Koppens2006}. However, options to tune the  initialisation rate or the targeted quantum state (e.g., selecting between a singlet or a triplet) are limited with these energy-based approaches.
On the other hand, exponential tuneability of QD tunneling rates, which has recently been shown to be highly effective \cite{Stein2017} for optimising the speed-precision trade-off of single-electron (SE) pumps in quantum metrology \cite{Kaestner2015}, has not yet been exploited to combine the speed and the selectivity in initialising individual orbital and/or spin states of electrons confined to QDs. Hence developing a strategy to control quantum-state-specific tunnelling 
into and out of a QD could open new ways for the efficient initialization of qubits and enable fast on-demand sources of specific few-particle quantum states 
for electron quantum optics  \cite{Bocquillon2012,Ubbelohde2015,Baeuerle2018} 
or the transfer of quantum information between static qubits \cite{Yamamoto2012,Baart2016natnano,Fujita2017,Flentje2017}.

In this work, we perform spectroscopy of quantum state energies and tunnel couplings for a dynamic QD operated as a SE pump by loading and capturing electrons near the Fermi edge.
When the lowest QD levels are close to the Fermi energy $\Ef$, incomplete loading of electrons can occur \cite{Leicht2010,Kaestner2015}
and the capture probability is determined by a competition of tunnel coupling, energy level positions \cite{Amasha2008} and loading times. 
We describe the experimental data by a simple model that enables quantitative spectroscopy for the rates of intunneling into the QD.
Measurements reveal that electrons can tunnel either directly or via excited orbital states into the SE ground state. In the latter case SE capture is mediated by fast orbital relaxation.
In contrast, relaxation for electron pair states requiring a singlet-triplet spin transition  occurs on a much longer timescale than the loading times. 
In this regime of electron pair capturing, our data clearly indicate loading into specific singlet or triplet states as function of gate voltages, thereby enabling gate controlled fast quantum state initialization of QDs.
\begin{figure}
  \includegraphics[width=0.98\columnwidth]{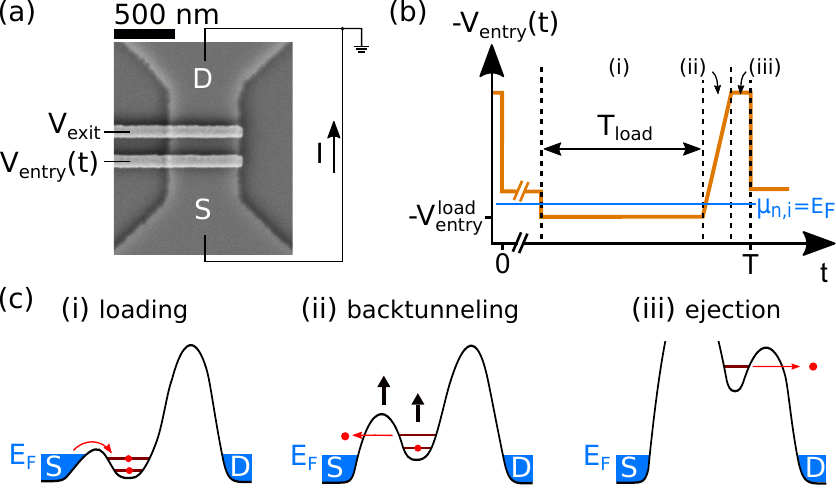}
  \caption{\label{fig:device}
	(a)~Scanning electron micrograph of a SE pump.
	(b)~A waveform with period $T=1/f$ is applied to the entry gate which moves the energy levels $\mu_{n,i}\propto -\Ventry(t)+\text{const}$.
	Loading of the QD is possible while $\mu_{n,i}<\Ef$.
	(c)~Measurement cycle:
	(i)~Electrons are loaded.
	(ii)~Backtunneling of higher energy electrons.
	(iii)~Captured electrons are ejected to drain.}
\end{figure}

Measurements were performed on a QD based on a GaAs/AlGaAs heterostructure as typically used for tunable barrier SE pumps \cite{Stein2015,Stein2017} [Fig.~\ref{fig:device}(a)].
From the two-dimensional electron system, a channel was formed by shallow wet etching and two Ti/Au finger-shaped Schottky gates were deposited on top.
At low temperatures ($T_{\mathrm{bath}}\approx100$\,mK), a QD is formed between entry and exit gate by applying negative voltages $\Ventry$ and $\Vexit$. Biased cooldown with +70~mV applied to both gates was used to increase device stability \cite{Pioro-Ladriere2005}.

The state of the QD is controlled by a time dependent  signal $\Ventry(t)$
with repetition frequency $f=1/T$ applied to the entry gate, while $\Vexit$ is kept constant. 
The shape of the clock signal [Fig.~\ref{fig:device}(b)] is designed to drive the QD through three distinct phases [Fig.~\ref{fig:device}(c)]:
(i):~For a time $\tload$, a number of lowest-energy states become available for  electrons to tunnel onto {an initially empty} QD.
(ii):~{The potential energy of the QD is raised,  and some electrons may escape back to the source as the entrance tunnel barrier is gradually pinched off. This allows separation of the electron states based on the difference in the backtunneling rates \cite{Kashcheyevs2010}. 
(iii):~All captured electrons are emitted to the drain through the exit barrier. The output current, $I=\langle n\rangle ef$, measures the average of the number $n$ of electrons captured  by the end of phase~(ii).

A large difference in energies and tunnel couplings between  one- and two-electron states is the basis for accurate tuning of the number of remaining electrons for quantum metrology of electrical current~\cite{Kaestner2015}.
In the example shown in Fig.~\ref{fig:oneemap}(a),  the voltage $\Vexit$ is used to control the backtunneling rates and hence the outcome of phase~(ii). Three plateaus of  quantized current corresponding to $\langle n \rangle =1,2,3$ are measured.
Adding $\VLE$ to the tuning parameters reveals a 2D map of $\langle n \rangle$ as function of $\VLE$ and $\Vexit$, shown in Fig.~\ref{fig:oneemap}(b).
$\VLE$ is the extreme value of $\Ventry(t)$, held constant for a time $\tload$ during the loading phase (i) [Fig.~\ref{fig:device}(b)].

We first consider the series of anti-diagonal steps highlighted by three blue lines in Fig.~\ref{fig:oneemap}(b).
We identify these lines as resonances of the source Fermi energy $\Ef$ and  the energies $\mu_{n,i}$  of distinct quantum states $i=0,1,2\ldots $ available for the first ($n=1$) electron.
If $\mu_{1,i}<\Ef$ during the loading phase, the state $i$ contributes to initializaiton with  
an intunneling rate $\gamma_{1,i}$. The values of both $\mu_{n,i}$ and $\gamma_{n,i}$ are tuned by the gate voltages, leading to qualitatively different initialization conditions sketched in Fig.\ref{fig:oneemap}(c).
\begin{figure}
  \includegraphics[width=0.98\columnwidth]{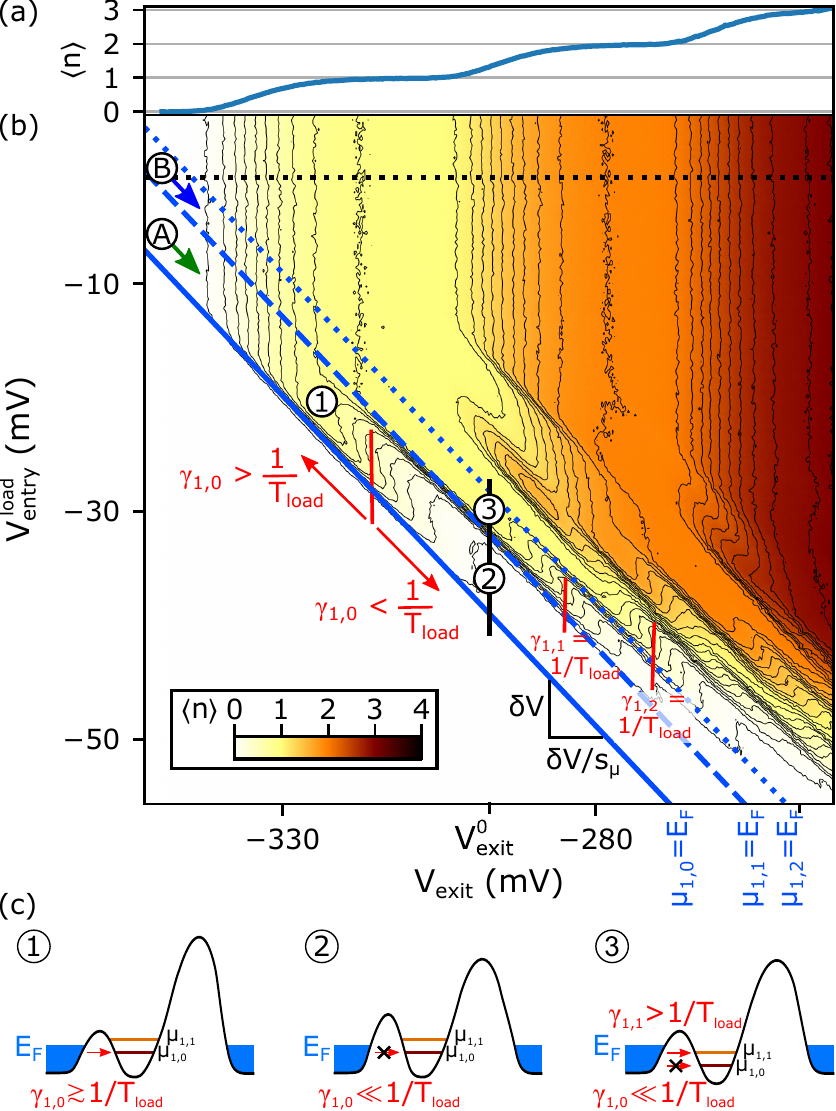}
  \caption{\label{fig:oneemap} 
	(a)~Quantized current steps are measured in the backtunneling-limited region  [along black dotted line in (b)].
	(b)~Measurement of the average number of captured electrons for $\tload=50$\,ns.
	(c)~Sketches of potential landscape during the loading phase (i), showing only the lowest two states.
	\textcircled{\scriptsize 1}~Only the lowest SE state $\mu_{1,0}$ is below $\Ef$ and can be loaded.
	\textcircled{\scriptsize 2}~The tunneling rate $\gamma_{1,0}$ is insufficient.
	\textcircled{\scriptsize 3}~Loading to the second state $\mu_{1,1}$ dominates.}
\end{figure}
In configuration \textcircled{\footnotesize 1}, only the ground state $\mu_{1,0}$ is  available for loading, whereas the excited state $\mu_{1,1}>\Ef$ is energetically forbidden. Here, an electron can only be loaded if $\gamma_{1,0}$ is sufficiently large compared to $1/\tload$. The corresponding values of control voltages ($\VLE$, $\Vexit$) are marked \textcircled{\footnotesize 1} in Fig.~\ref{fig:oneemap}(b). Configuration
\textcircled{\footnotesize 2} is energetically similar to configuration \textcircled{\footnotesize 1},
but the entry barrier is higher, see  Fig.\ref{fig:oneemap}(c),  and thus the intunneling rate of the only energetically available state is much lower, $\gamma_{1,0}\ll1/\tload$. Hence loading at \textcircled{\footnotesize 2} fails (negligible $\langle n \rangle$).

In configuration~\textcircled{\footnotesize 3} a second level is energetically available and can contribute to loading.
This higher energy state $i\!=\!1$ is separated from the source by an effectively lower tunnel barrier resulting in an exponentially stronger coupling compared to the ground state $i\!=\!0$.
In this regime $\gamma_{1,0}\ll1/\tload<\gamma_{1,1}$, and the loading takes place predominantly via state $i=1$. 

We develop the above qualitative picture into a quantitative model for tunnelling rate spectroscopy.
{A sufficiently sharp Fermi edge ensures that the processes of tunneling in (i) and  out (ii) of the dot [see Fig 1(b)] are temporally separated.
Hence the  current at 
a given point ($\VLE$, $\Vexit$) 
near the lower-left corner of the {$\langle n \rangle =1$} plateau can be modelled
by a product of two probabilities, $P_{\text{load}}$ for loading and $1-P_{\text{back}}$ for not backtunneling, respectively:
}
\begin{equation}
  \langle{n}\rangle=P_{\mathrm{load}}\cdot(1-P_{\mathrm{back}}).
\label{eq:I}
\end{equation}
Extracting $P_{\text{load}} = 1-\exp[-\Gamma_{\text{in}}( \VLE, \Vexit) \, \tload]$ gives direct information on the total intunnelling rate
$\Gamma_{\mathrm{in}}$ during the loading stage.
The other probability, $P_{\text{back}} =1-\exp{\{ -\int_{} \Gamma_{\text{back}}[ \Ventry(t), \Vexit]} dt \} $, 
depends on the electron escape rate $\Gamma_{\text{back}}$ during the backtunneling phase (ii),
and can be parametrised as $P_{\mathrm{back}}=1-\exp(-e^{-\alpha \Vexit+\delta_{\text{back}}})$ \cite{Kashcheyevs2010,Fricke2013,Yamahata2014a,Rossi2018}.
Both rates are defined by the same tunnel barrier, hence we expect the same exponential parametric dependence on the gate voltages for the in- and backtunneling rates, $\Gamma_{\text{in}} , \Gamma_{\text{back}} \propto e^{-\alpha \Vexit }$~\cite{MacLean2007}. This is valid 
along the lines of constant energy, 
identified by a common slope $s_\mu$ as $\mu_{n,i}(\Ventry,\Vexit) = -k_\mu (\Ventry +\!s_{\mu} \Vexit) +\text{const}$ in Fig.~\ref{fig:oneemap}(b) ($k_\mu$ is the measure of the gate lever arm; $s_{\mu}>0$). For DC voltage shifts along these lines, the time-evolution of the QD energies is not affected, and hence neither the number of discrete states competing for loading, nor the timing for the onset of backtunneling~\cite{Fricke2013,Rossi2018} change. 

Technically, we use the following ansatz to describe the gate voltage dependence of the intunneling rate:
\begin{align}
	\Gamma_{\mathrm{in}}\left( \VLE, \Vexit \right) = W_{\mathrm{in}}\left(\VLE \!+\!s_{\mu} \Delta \Vexit \right) \, e^{-\alpha \Delta \Vexit }, \label{eq:gammain}
\end{align}
{where $\Delta \Vexit = \Vexit -\Vexit^{0}$ is measured from an arbitrary chosen reference level $\Vexit^{0}=-297 \, \text{mV}$,
and an unknown single-variable function $W_{\mathrm{in}}$ 
describes the dependency of the rate on the depth of the QD during the loading stage. 
The function $W_{\mathrm{in}}\left(V\right)$ is deduced by fitting Eq.~\eqref{eq:I} along the lines of constant $V=\VLE \!+\!s_{\mu} \Delta \Vexit$ with $V$-independent globally optimised values of $s_{\mu}$ and $\alpha$~\cite{supp}. Two representative fits are shown in Figs.~\ref{fig:fits}(a),~(b), differing only in the best-fit value of $W_{\text{in}}$.
}
\begin{figure}
	\includegraphics[width=0.98\columnwidth]{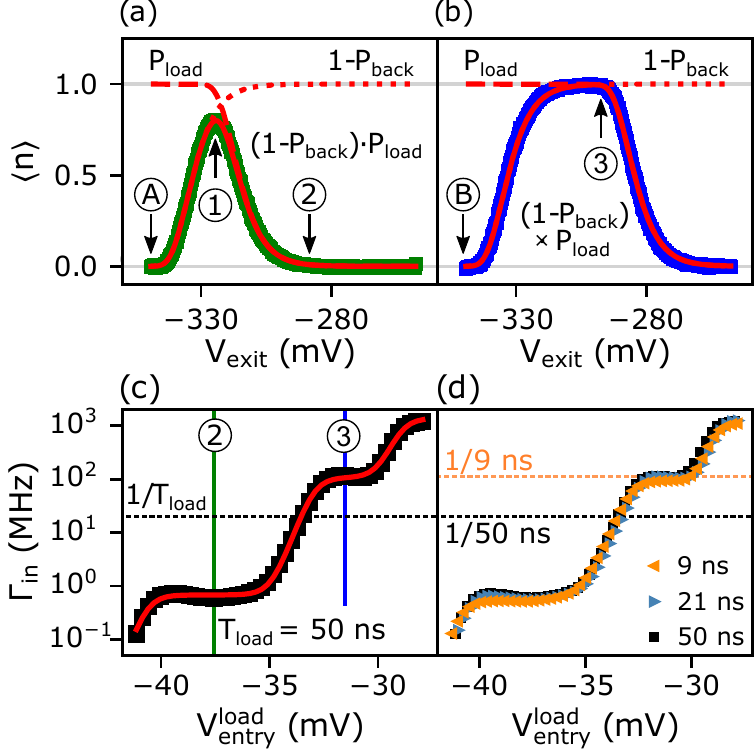}
	\caption{
	(a,b)~Electron number $\langle n \rangle$ measured along a line of constant energy, starting at point~\textcircled{\scriptsize A} and point~\textcircled{\scriptsize B} in Fig.~\ref{fig:oneemap}(b), respectively.
	Red lines indicate fits using Eq.~(\ref{eq:I}).
	(c,d)~Tunneling rates $\Gamma_{\mathrm{in}}$ for fixed $\Vexit=\Vexit^{0}$ determined from fits for various values of $\VLE$ along the black line in Fig.\ref{fig:oneemap}(b) for (c) $\tload=9\,$ns and (d) $\tload=9, 21$, and $50\,$ns.
	Red line in (c) indicates a fit using Eq.~(\ref{eq:gamma}).}
	\label{fig:fits}
\end{figure}

{The method measures $\Gamma_{\text{in}}$ directly only when $\Gamma_{\mathrm{in}}\left( \VLE, \Vexit \right)$ is on the same order of magnitude as $1/T_{\text{load}}$ [regions near the red lines $\gamma_{1,i} =1/\tload$ in Fig.~\ref{fig:oneemap}(b)], and
relies on extrapolation of the exponential parametrization  to extend the gate-voltage dependence via Eq.~\eqref{eq:gammain} to the whole SE loading region ($\Ef < \mu_{2,0}$).}
The resulting $\Gamma_{\mathrm{in}}\left( \VLE, \Vexit^0 \right)$ is shown in Fig.~\ref{fig:fits}(c).

An important consistency check is independence of the inferred $\Gamma_{\text{in}}$ on the chosen value of $\tload$. Measurements with $\tload=9$~and $21$~ns result in reduced loading probabilities compared to $\tload=50$~ns, but the extracted
$\Gamma_{\mathrm{in}}$ values agree well, as shown in Fig.~\ref{fig:fits}(d).
 
Figures~\ref{fig:fits}(c) and (d) show three clear steps of increased tunnel coupling  corresponding to the (blue) constant-energy lines in Fig.~\ref{fig:oneemap}(b). 
{We parametrize the total rate by a discrete sum,} 
\begin{equation}
	W_{\mathrm{in}}(\VLE) = \sum_{i=0}^{2} \gamma_{1,i}^{0} f\left(\VLE-\Ventry^{1,i}\right) \, ,
\label{eq:gamma}
\end{equation}
where $f(V)=[\exp(-k_\mu V/k_{\mathrm{B}}{\mathrm{T}}_{\text{eff}})+1]^{-1}$ is the Fermi distribution,  $k_{\mathrm{B}}$ is the Boltzmann constant, and $\Ventry^{1,i}$ is the value of $\VLE$ at which $\mu_{1,i}(\VLE, \Vexit^0)=\Ef$.
Equation~\eqref{eq:gamma} assumes that each state $\mu_{1,i}$ contributes to the total rate $\Gamma_{\mathrm{in}}$ with $\gamma_{1,i}=\gamma_{1,i}^{0}\, e^{-\alpha \Delta \Vexit } $ when it is energetically possible according to the schematics of Fig.~\ref{fig:oneemap}(c). Here $\mathrm{T}_{\text{eff}}$ is the effective temperature and $\mathrm{T}_{\text{eff}}/k_\mu$, $\Ventry^{1,i}$ and $\gamma_{1,i}^{0}$ are used as fitting parameters~\cite{supp}.
In Figure~\ref{fig:fits}(c) the fit to Eq.~\eqref{eq:gamma} (red line) describes the data (squares) well.
We find rather large differences of tunneling rates for consecutive SE states, $\gamma_{1,1}/\gamma_{1,0}\approx 160$ and $\gamma_{1,2}/\gamma_{1,1}\approx 12$.

In contrast to  clear signatures of distinct SE states during the loading phase, the backtunneling probability,  also inferred from fits to Eq.~\eqref{eq:I}, shows no appreciable dependence on $\VLE$.
This implies that by the end of loading the electron has relaxed to a state with a unique backtunnelling rate, independent of the intunneling channel $i$.
This is consistent with the expectation of strong relaxation from the excited orbital states ($i>0$) to the SE ground state ($i=0$) on a time scale that is much faster than the loading times of our experiment \cite{Stano2006}.

We further clarify the roles of relaxation  by comparing several  theoretical scenarios, all sharing the same  gate-voltage dependence of individual tunnelling rates  inferred from the experiment via Eq.~\eqref{eq:gamma}.
In Fig.~\ref{fig:twoemaps}(a) we simulate the detector signal with the ground-state level $i\!=\!0$ only. Here, electrons capture succeeds only in a narrow parameter range, limited on the right by incomplete loading  and
on the left by backtunneling.  In Fig.~\ref{fig:twoemaps}(b) we include 
three states, all quickly relaxing to $i\!=\!0$. The model accurately matches the 
experimental data in the relevant range [cf.~Fig.~\ref{fig:twoemaps}(d)] as expected from the robustness of the fits in Fig.~\ref{fig:fits}. 

A hypothetical SE capture scenario in Fig.~\ref{fig:twoemaps}(c) simulates the same three states as in (b) but without internal relaxation~\cite{supp}. The state-specific backtunneling rates $\gamma_{n,i}^{\text{back}}$ are derived from $\gamma_{n,i}^{}=g_{n,i}^{} \, \gamma_{n,i}^{\text{back}}$ where 
$g_{n,i}$ is the degeneracy factor~\cite{MacLean2007,Rossi2018} ($g_{1,i}=2$ for spin).
In sharp contrast to (b), scenario (c) presents a gap in the current (region G1 in the figure) which can be seen to arise from a Coulomb blockade by a better-coupled state~\cite{Rossi2018}: for example, as soon as the excited SE state (1,1) becomes energetically available at the start of the loading phase (i), it  gets immediately occupied due to much larger tunnel coupling, thus blocking further loading into the  ground state (1,0) due to Coulomb repulsion. In this scenario without relaxation, the occupied (1,1) state  decays back to the source as soon as this becomes energetically possible in phase (ii), resulting in the gap G1. }

\begin{figure}
	\includegraphics[width=\columnwidth]{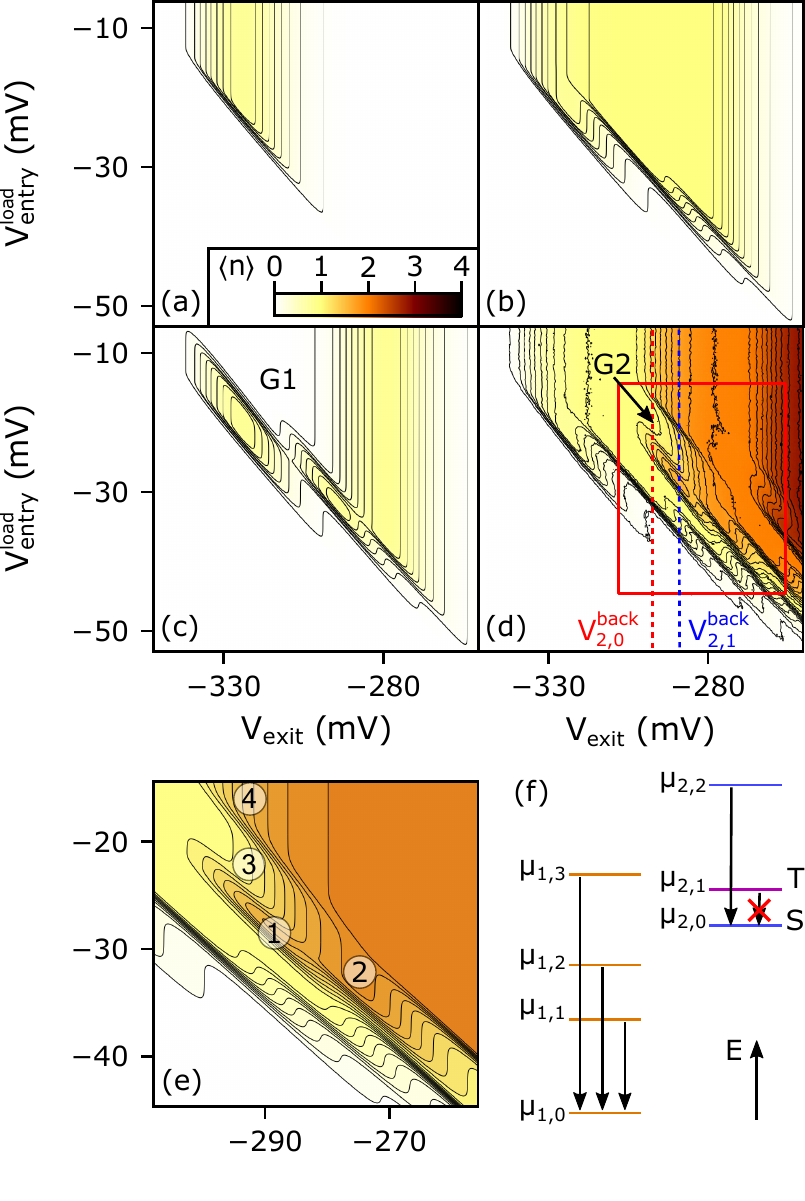} 
	\caption{
	(a,b,c) Simulations of the initialized electron number $\langle n \rangle$  for three models:
	(a)~the SE ground state $i\!=\!0$ alone,
	(b)~three SE states ($i\!=\!0,1,2$) with immediate relaxation, and
	(c)~three SE states ($i\!=\!0,1,2$) without relaxation.
	(d)~Measured $\langle n \rangle$ for the same gate voltage range.
	(e)~Simulation with multiple single- and two-electron states with energies
	and relaxation paths depicted in (f). The corresponding region is highlighted by a red box in (d).}
	\label{fig:twoemaps}
\end{figure}

Turning now to the experimental data on electron pairs, we note a similar gap [G2 in Fig.~\ref{fig:twoemaps}(d)] in the $n=2$ electron backtunneling line, shifting the latter from  $\Vexit=V_{\mathrm{back}}^{2,0}$ towards more positive $\Vexit=V_{\mathrm{back}}^{2,1}$. This suggests initialization of an excited state (2,1) which persists unrelaxed, and which has a backtunneling rate $\gamma^{\text{back}}_{2,1}/\gamma^{\text{back}}_{2,0}\approx \exp[\alpha(V_{\mathrm{back}}^{2,1}- V_{\mathrm{back}}^{2,0})] \approx 5$ times higher than the ground state (2,0). 
Extending the model to include loading and backtunneling of a second electron at discrete energies $\mu_{2,i}$ with appropriately chosen rates $\gamma_{2,i}$ and blocked $(2,1)\to(2,0)$ relaxation enables the simulation [Fig.~\ref{fig:twoemaps}(e)] to match quite well the structure of the observed energy-, intunneling-, and backtunneling-limited lines [cf.\ Fig.~\ref{fig:twoemaps}(d)]. 
The relative energies and relaxation paths assigned to the simulation are shown in Fig.~\ref{fig:twoemaps}(f), quantitative details are given in the Supplemental Material~\cite{supp}.

The observed suppression of relaxation can be explained naturally by identifying $(2,0)$ and $(2,1)$ with  singlet (S) and triplet (T) states, respectively, because triplet-to-singlet relaxation
requires a spin-flip \cite{Hanson2007} which occurs on a timescale typically  much longer than our range of $\tload$ \cite{Fujisawa2002,Sasaki2005,Hanson2005}.

The key regions demonstrating control of electron pair initialization are marked in Fig.~\ref{fig:twoemaps}(e)  as $\textcircled{\footnotesize 1}$-$\textcircled{\footnotesize 4}$.
$\textcircled{\footnotesize 1}$: Only ground-state $S$ intialization is energetically allowed.
$\textcircled{\footnotesize 2}$ and $\textcircled{\footnotesize 3}$: Loading into both T and S is allowed energetically but T is dynamically preferred due to 
a larger tunnelling rate. The selectively initialized T state is kept at $\textcircled{\footnotesize 2}$ but lost at $\textcircled{\footnotesize 3}$ due to backtunnelling.
$\textcircled{\footnotesize 4}$: Higher-energy S-type states enter, resulting in pre-dominantly ground-state S initialisation  via quick orbital relaxation paths not requiring a spin flip.

Switching the DC voltages between points \textcircled{\footnotesize 1} and \textcircled{\footnotesize 2} allows generation of electron pairs with on-demand selection between S and T states. The probability of capturing  a pair instead of a SE can be estimated  as $\langle n \rangle-1$. The corresponding maximal values for $\tload =50$~ns are $55\%$ and $80\%$, for on-demand S and T respectively; both values are straightforward to improve by increasing $\tload$. The fidelity of preparation  for any of the three components ($g_{2,1}=3$) of the excited state T  versus the ground state S ($g_{2,0}=1$)  at \textcircled{\footnotesize 2} can be approximated by the ratio of the corresponding intunnelling rates, $\gamma_{2,1}/ \gamma_{2,0} \approx 15$.
These values are dictated by the energy gap and the barrier selectivity inherent to particular device; 
we envision the fidelity can be further improved by optimizing the confinement potential of the QD.

{In summary, we have presented a method to identify, tune and measure tunelling rates for discrete single- and two-electron quantum states of a semiconductor QD. Exponential energy-dependance of the individual rates enable selective initialisation of quantum states on the time scales shorter than spin-relaxation time.
Our device can be used as an electron ``entangler'' for on-demand emission of electron pairs with deterministically controlled exchange symmetry for applications in electron quantum optics \cite{Fletcher2013,Ubbelohde2015,Baeuerle2018} and quantum information transfer \cite{Fujita2017,Flentje2017}.
}

\begin{acknowledgments}
This work was supported in part by the Joint Research Projects e-SI-Amp (15SIB08) and SEQUOIA (17FUN04). 
These projects received funding from the European Metrology Programme for Innovation and Research (EMPIR) co-financed by the Participating States and from the European Unions Horizon 2020 research and innovation programme. J.~K. has been supported by University
of Latvia grant no.~AAP2016/B031. We thank Holger Marx and Klaus Pierz for MBE growth, Thomas Weimann for electron-beam lithography, and Rolf J. Haug for stimulating discussions.
\end{acknowledgments}


\begin{thebibliography}{30}%
	\makeatletter
	\providecommand \@ifxundefined [1]{%
		\@ifx{#1\undefined}
	}%
	\providecommand \@ifnum [1]{%
		\ifnum #1\expandafter \@firstoftwo
		\else \expandafter \@secondoftwo
		\fi
	}%
	\providecommand \@ifx [1]{%
		\ifx #1\expandafter \@firstoftwo
		\else \expandafter \@secondoftwo
		\fi
	}%
	\providecommand \natexlab [1]{#1}%
	\providecommand \enquote  [1]{``#1''}%
	\providecommand \bibnamefont  [1]{#1}%
	\providecommand \bibfnamefont [1]{#1}%
	\providecommand \citenamefont [1]{#1}%
	\providecommand \href@noop [0]{\@secondoftwo}%
	\providecommand \href [0]{\begingroup \@sanitize@url \@href}%
	\providecommand \@href[1]{\@@startlink{#1}\@@href}%
	\providecommand \@@href[1]{\endgroup#1\@@endlink}%
	\providecommand \@sanitize@url [0]{\catcode `\\12\catcode `\$12\catcode
		`\&12\catcode `\#12\catcode `\^12\catcode `\_12\catcode `\%12\relax}%
	\providecommand \@@startlink[1]{}%
	\providecommand \@@endlink[0]{}%
	\providecommand \url  [0]{\begingroup\@sanitize@url \@url }%
	\providecommand \@url [1]{\endgroup\@href {#1}{\urlprefix }}%
	\providecommand \urlprefix  [0]{URL }%
	\providecommand \Eprint [0]{\href }%
	\providecommand \doibase [0]{https://doi.org/}%
	\providecommand \selectlanguage [0]{\@gobble}%
	\providecommand \bibinfo  [0]{\@secondoftwo}%
	\providecommand \bibfield  [0]{\@secondoftwo}%
	\providecommand \translation [1]{[#1]}%
	\providecommand \BibitemOpen [0]{}%
	\providecommand \bibitemStop [0]{}%
	\providecommand \bibitemNoStop [0]{.\EOS\space}%
	\providecommand \EOS [0]{\spacefactor3000\relax}%
	\providecommand \BibitemShut  [1]{\csname bibitem#1\endcsname}%
	\let\auto@bib@innerbib\@empty
	\bibitem [{\citenamefont {DiVincenzo}(2000)}]{DiVincenzo2000}%
	\BibitemOpen
	\bibfield  {author} {\bibinfo {author} {\bibfnamefont {D.~P.}\ \bibnamefont
			{DiVincenzo}},\ }\bibfield  {title} {\bibinfo {title} {The physical
			implementation of quantum computation},\ }\href@noop {} {\bibfield  {journal}
		{\bibinfo  {journal} {Fortschritte der Physik}\ }\textbf {\bibinfo {volume}
			{48}},\ \bibinfo {pages} {771} (\bibinfo {year} {2000})}\BibitemShut
	{NoStop}%
	\bibitem [{\citenamefont {Knill}(2005)}]{Knill2005}%
	\BibitemOpen
	\bibfield  {author} {\bibinfo {author} {\bibfnamefont {E.}~\bibnamefont
			{Knill}},\ }\bibfield  {title} {\bibinfo {title} {{Quantum computing with
				realistically noisy devices}},\ }\href {https://doi.org/10.1038/nature03350}
	{\bibfield  {journal} {\bibinfo  {journal} {Nature}\ }\textbf {\bibinfo
			{volume} {434}},\ \bibinfo {pages} {39} (\bibinfo {year} {2005})}\BibitemShut
	{NoStop}%
	\bibitem [{\citenamefont {Loss}\ and\ \citenamefont
		{DiVincenzo}(1998)}]{Loss1998}%
	\BibitemOpen
	\bibfield  {author} {\bibinfo {author} {\bibfnamefont {D.}~\bibnamefont
			{Loss}}\ and\ \bibinfo {author} {\bibfnamefont {D.~P.}\ \bibnamefont
			{DiVincenzo}},\ }\bibfield  {title} {\bibinfo {title} {Quantum computation
			with quantum dots},\ }\href {https://doi.org/10.1103/PhysRevA.57.120}
	{\bibfield  {journal} {\bibinfo  {journal} {Phys. Rev. A}\ }\textbf {\bibinfo
			{volume} {57}},\ \bibinfo {pages} {120} (\bibinfo {year} {1998})}\BibitemShut
	{NoStop}%
	\bibitem [{\citenamefont {Petta}\ \emph {et~al.}(2005)\citenamefont {Petta},
		\citenamefont {Johnson}, \citenamefont {Taylor}, \citenamefont {Laird},
		\citenamefont {Yacoby}, \citenamefont {Lukin}, \citenamefont {Marcus},
		\citenamefont {Hanson},\ and\ \citenamefont {Gossard}}]{Petta2005}%
	\BibitemOpen
	\bibfield  {author} {\bibinfo {author} {\bibfnamefont {J.~R.}\ \bibnamefont
			{Petta}}, \bibinfo {author} {\bibfnamefont {A.~C.}\ \bibnamefont {Johnson}},
		\bibinfo {author} {\bibfnamefont {J.~M.}\ \bibnamefont {Taylor}}, \bibinfo
		{author} {\bibfnamefont {E.~A.}\ \bibnamefont {Laird}}, \bibinfo {author}
		{\bibfnamefont {A.}~\bibnamefont {Yacoby}}, \bibinfo {author} {\bibfnamefont
			{M.~D.}\ \bibnamefont {Lukin}}, \bibinfo {author} {\bibfnamefont {C.~M.}\
			\bibnamefont {Marcus}}, \bibinfo {author} {\bibfnamefont {M.~P.}\
			\bibnamefont {Hanson}},\ and\ \bibinfo {author} {\bibfnamefont {A.~C.}\
			\bibnamefont {Gossard}},\ }\bibfield  {title} {\bibinfo {title} {{Coherent
				Manipulation of Coupled Electron Spins in Semiconductor Quantum Dots}},\
	}\href {https://doi.org/10.1126/science.1116955} {\bibfield  {journal}
		{\bibinfo  {journal} {Science}\ }\textbf {\bibinfo {volume} {309}},\ \bibinfo
		{pages} {2180} (\bibinfo {year} {2005})}\BibitemShut {NoStop}%
	\bibitem [{\citenamefont {Koppens}\ \emph {et~al.}(2006)\citenamefont
		{Koppens}, \citenamefont {Buizert}, \citenamefont {Tielrooij}, \citenamefont
		{Vink}, \citenamefont {Nowack}, \citenamefont {Meunier}, \citenamefont
		{Kouwenhoven},\ and\ \citenamefont {Vandersypen}}]{Koppens2006}%
	\BibitemOpen
	\bibfield  {author} {\bibinfo {author} {\bibfnamefont {F.~H.~L.}\
			\bibnamefont {Koppens}}, \bibinfo {author} {\bibfnamefont {C.}~\bibnamefont
			{Buizert}}, \bibinfo {author} {\bibfnamefont {K.~J.}\ \bibnamefont
			{Tielrooij}}, \bibinfo {author} {\bibfnamefont {I.~T.}\ \bibnamefont {Vink}},
		\bibinfo {author} {\bibfnamefont {K.~C.}\ \bibnamefont {Nowack}}, \bibinfo
		{author} {\bibfnamefont {T.}~\bibnamefont {Meunier}}, \bibinfo {author}
		{\bibfnamefont {L.~P.}\ \bibnamefont {Kouwenhoven}},\ and\ \bibinfo {author}
		{\bibfnamefont {L.~M.~K.}\ \bibnamefont {Vandersypen}},\ }\bibfield  {title}
	{\bibinfo {title} {{Driven coherent oscillations of a single electron spin in
				a quantum dot}},\ }\href {https://doi.org/10.1038/nature05065} {\bibfield
		{journal} {\bibinfo  {journal} {Nature}\ }\textbf {\bibinfo {volume} {442}},\
		\bibinfo {pages} {766} (\bibinfo {year} {2006})}\BibitemShut {NoStop}%
	\bibitem [{\citenamefont {Stein}\ \emph {et~al.}(2017)\citenamefont {Stein},
		\citenamefont {Scherer}, \citenamefont {Gerster}, \citenamefont {Behr},
		\citenamefont {G\"otz}, \citenamefont {Pesel}, \citenamefont {Leicht},
		\citenamefont {Ubbelohde}, \citenamefont {Weimann}, \citenamefont {Pierz},
		\citenamefont {Schumacher},\ and\ \citenamefont {Hohls}}]{Stein2017}%
	\BibitemOpen
	\bibfield  {author} {\bibinfo {author} {\bibfnamefont {F.}~\bibnamefont
			{Stein}}, \bibinfo {author} {\bibfnamefont {H.}~\bibnamefont {Scherer}},
		\bibinfo {author} {\bibfnamefont {T.}~\bibnamefont {Gerster}}, \bibinfo
		{author} {\bibfnamefont {R.}~\bibnamefont {Behr}}, \bibinfo {author}
		{\bibfnamefont {M.}~\bibnamefont {G\"otz}}, \bibinfo {author} {\bibfnamefont
			{E.}~\bibnamefont {Pesel}}, \bibinfo {author} {\bibfnamefont
			{C.}~\bibnamefont {Leicht}}, \bibinfo {author} {\bibfnamefont
			{N.}~\bibnamefont {Ubbelohde}}, \bibinfo {author} {\bibfnamefont
			{T.}~\bibnamefont {Weimann}}, \bibinfo {author} {\bibfnamefont
			{K.}~\bibnamefont {Pierz}}, \bibinfo {author} {\bibfnamefont {H.~W.}\
			\bibnamefont {Schumacher}},\ and\ \bibinfo {author} {\bibfnamefont
			{F.}~\bibnamefont {Hohls}},\ }\bibfield  {title} {\bibinfo {title}
		{Robustness of single-electron pumps at sub-ppm current accuracy level},\
	}\href {https://doi.org/10.1088/1681-7575/54/1/S1} {\bibfield  {journal}
		{\bibinfo  {journal} {Metrologia}\ }\textbf {\bibinfo {volume} {54}},\
		\bibinfo {pages} {S1} (\bibinfo {year} {2017})}\BibitemShut {NoStop}%
	\bibitem [{\citenamefont {Kaestner}\ and\ \citenamefont
		{Kashcheyevs}(2015)}]{Kaestner2015}%
	\BibitemOpen
	\bibfield  {author} {\bibinfo {author} {\bibfnamefont {B.}~\bibnamefont
			{Kaestner}}\ and\ \bibinfo {author} {\bibfnamefont {V.}~\bibnamefont
			{Kashcheyevs}},\ }\bibfield  {title} {\bibinfo {title} {Non-adiabatic
			quantized charge pumping with tunable-barrier quantum dots: a review of
			current progress},\ }\href {https://doi.org/10.1088/0034-4885/78/10/103901}
	{\bibfield  {journal} {\bibinfo  {journal} {Rep. Prog. Phys.}\ }\textbf
		{\bibinfo {volume} {78}},\ \bibinfo {pages} {103901} (\bibinfo {year}
		{2015})}\BibitemShut {NoStop}%
	\bibitem [{\citenamefont {Bocquillon}\ \emph {et~al.}(2012)\citenamefont
		{Bocquillon}, \citenamefont {Parmentier}, \citenamefont {Grenier},
		\citenamefont {Berroir}, \citenamefont {Degiovanni}, \citenamefont {Glattli},
		\citenamefont {Pla\c{c}ais}, \citenamefont
		{Cavanna}, \citenamefont {Jin},\ and\ \citenamefont
		{F\`eve}}]{Bocquillon2012}%
	\BibitemOpen
	\bibfield  {author} {\bibinfo {author} {\bibfnamefont {E.}~\bibnamefont
			{Bocquillon}}, \bibinfo {author} {\bibfnamefont {F.~D.}\ \bibnamefont
			{Parmentier}}, \bibinfo {author} {\bibfnamefont {C.}~\bibnamefont {Grenier}},
		\bibinfo {author} {\bibfnamefont {J.-M.}\ \bibnamefont {Berroir}}, \bibinfo
		{author} {\bibfnamefont {P.}~\bibnamefont {Degiovanni}}, \bibinfo {author}
		{\bibfnamefont {D.~C.}\ \bibnamefont {Glattli}}, \bibinfo {author}
		{\bibfnamefont {B.}~\bibnamefont {Pla\ifmmode~\mbox{\c{c}}\else
				\c{c}\fi{}ais}}, \bibinfo {author} {\bibfnamefont {A.}~\bibnamefont
			{Cavanna}}, \bibinfo {author} {\bibfnamefont {Y.}~\bibnamefont {Jin}},\ and\
		\bibinfo {author} {\bibfnamefont {G.}~\bibnamefont {F\`eve}},\ }\bibfield
	{title} {\bibinfo {title} {Electron quantum optics: Partitioning electrons
			one by one},\ }\href {https://doi.org/10.1103/PhysRevLett.108.196803}
	{\bibfield  {journal} {\bibinfo  {journal} {Phys. Rev. Lett.}\ }\textbf
		{\bibinfo {volume} {108}},\ \bibinfo {pages} {196803} (\bibinfo {year}
		{2012})}\BibitemShut {NoStop}%
	\bibitem [{\citenamefont {Ubbelohde}\ \emph {et~al.}(2015)\citenamefont
		{Ubbelohde}, \citenamefont {Hohls}, \citenamefont {Kashcheyevs},
		\citenamefont {Wagner}, \citenamefont {Fricke}, \citenamefont {K\"astner},
		\citenamefont {Pierz}, \citenamefont {Schumacher},\ and\ \citenamefont
		{Haug}}]{Ubbelohde2015}%
	\BibitemOpen
	\bibfield  {author} {\bibinfo {author} {\bibfnamefont {N.}~\bibnamefont
			{Ubbelohde}}, \bibinfo {author} {\bibfnamefont {F.}~\bibnamefont {Hohls}},
		\bibinfo {author} {\bibfnamefont {V.}~\bibnamefont {Kashcheyevs}}, \bibinfo
		{author} {\bibfnamefont {T.}~\bibnamefont {Wagner}}, \bibinfo {author}
		{\bibfnamefont {L.}~\bibnamefont {Fricke}}, \bibinfo {author} {\bibfnamefont
			{B.}~\bibnamefont {K\"astner}}, \bibinfo {author} {\bibfnamefont
			{K.}~\bibnamefont {Pierz}}, \bibinfo {author} {\bibfnamefont {H.~W.}\
			\bibnamefont {Schumacher}},\ and\ \bibinfo {author} {\bibfnamefont {R.~J.}\
			\bibnamefont {Haug}},\ }\bibfield  {title} {\bibinfo {title} {Partitioning of
			on-demand electron pairs},\ }\href {https://doi.org/10.1038/nnano.2014.275}
	{\bibfield  {journal} {\bibinfo  {journal} {Nature Nanotechnology}\ }\textbf
		{\bibinfo {volume} {10}},\ \bibinfo {pages} {46} (\bibinfo {year}
		{2015})}\BibitemShut {NoStop}%
	\bibitem [{\citenamefont {B\"auerle}\ \emph {et~al.}(2018)\citenamefont
		{B\"auerle}, \citenamefont {Glattli}, \citenamefont {Meunier}, \citenamefont
		{Portier}, \citenamefont {Roche}, \citenamefont {Roulleau}, \citenamefont
		{Takada},\ and\ \citenamefont {Waintal}}]{Baeuerle2018}%
	\BibitemOpen
	\bibfield  {author} {\bibinfo {author} {\bibfnamefont {C.}~\bibnamefont
			{B\"auerle}}, \bibinfo {author} {\bibfnamefont {D.~C.}\ \bibnamefont
			{Glattli}}, \bibinfo {author} {\bibfnamefont {T.}~\bibnamefont {Meunier}},
		\bibinfo {author} {\bibfnamefont {F.}~\bibnamefont {Portier}}, \bibinfo
		{author} {\bibfnamefont {P.}~\bibnamefont {Roche}}, \bibinfo {author}
		{\bibfnamefont {P.}~\bibnamefont {Roulleau}}, \bibinfo {author}
		{\bibfnamefont {S.}~\bibnamefont {Takada}},\ and\ \bibinfo {author}
		{\bibfnamefont {X.}~\bibnamefont {Waintal}},\ }\bibfield  {title} {\bibinfo
		{title} {Coherent control of single electrons: a review of current
			progress},\ }\href {https://doi.org/10.1088/1361-6633/aaa98a} {\bibfield
		{journal} {\bibinfo  {journal} {Rep. Prog. Phys.}\ }\textbf {\bibinfo
			{volume} {81}},\ \bibinfo {pages} {056503} (\bibinfo {year}
		{2018})}\BibitemShut {NoStop}%
	\bibitem [{\citenamefont {Yamamoto}\ \emph {et~al.}(2012)\citenamefont
		{Yamamoto}, \citenamefont {Takada}, \citenamefont {Bäuerle}, \citenamefont
		{Watanabe}, \citenamefont {Wieck},\ and\ \citenamefont
		{Tarucha}}]{Yamamoto2012}%
	\BibitemOpen
	\bibfield  {author} {\bibinfo {author} {\bibfnamefont {M.}~\bibnamefont
			{Yamamoto}}, \bibinfo {author} {\bibfnamefont {S.}~\bibnamefont {Takada}},
		\bibinfo {author} {\bibfnamefont {C.}~\bibnamefont {Bäuerle}}, \bibinfo
		{author} {\bibfnamefont {K.}~\bibnamefont {Watanabe}}, \bibinfo {author}
		{\bibfnamefont {A.~D.}\ \bibnamefont {Wieck}},\ and\ \bibinfo {author}
		{\bibfnamefont {S.}~\bibnamefont {Tarucha}},\ }\bibfield  {title} {\bibinfo
		{title} {Electrical control of a solid-state flying qubit},\ }\href
	{https://doi.org/10.1038/nnano.2012.28} {\bibfield  {journal} {\bibinfo
			{journal} {Nature Nanotechnology}\ }\textbf {\bibinfo {volume} {7}},\
		\bibinfo {pages} {247} (\bibinfo {year} {2012})}\BibitemShut {NoStop}%
	\bibitem [{\citenamefont {Baart}\ \emph {et~al.}(2016)\citenamefont {Baart},
		\citenamefont {Shafiei}, \citenamefont {Fujita}, \citenamefont {Reichl},
		\citenamefont {Wegscheider},\ and\ \citenamefont
		{Vandersypen}}]{Baart2016natnano}%
	\BibitemOpen
	\bibfield  {author} {\bibinfo {author} {\bibfnamefont {T.~A.}\ \bibnamefont
			{Baart}}, \bibinfo {author} {\bibfnamefont {M.}~\bibnamefont {Shafiei}},
		\bibinfo {author} {\bibfnamefont {T.}~\bibnamefont {Fujita}}, \bibinfo
		{author} {\bibfnamefont {C.}~\bibnamefont {Reichl}}, \bibinfo {author}
		{\bibfnamefont {W.}~\bibnamefont {Wegscheider}},\ and\ \bibinfo {author}
		{\bibfnamefont {L.~M.~K.}\ \bibnamefont {Vandersypen}},\ }\bibfield  {title}
	{\bibinfo {title} {{Single-spin CCD}},\ }\href
	{https://doi.org/10.1038/nnano.2015.291} {\bibfield  {journal} {\bibinfo
			{journal} {Nature Nanotechnology}\ }\textbf {\bibinfo {volume} {11}},\
		\bibinfo {pages} {330} (\bibinfo {year} {2016})}\BibitemShut {NoStop}%
	\bibitem [{\citenamefont {Fujita}\ \emph {et~al.}(2017)\citenamefont {Fujita},
		\citenamefont {Baart}, \citenamefont {Reichl}, \citenamefont {Wegscheider},\
		and\ \citenamefont {Vandersypen}}]{Fujita2017}%
	\BibitemOpen
	\bibfield  {author} {\bibinfo {author} {\bibfnamefont {T.}~\bibnamefont
			{Fujita}}, \bibinfo {author} {\bibfnamefont {T.~A.}\ \bibnamefont {Baart}},
		\bibinfo {author} {\bibfnamefont {C.}~\bibnamefont {Reichl}}, \bibinfo
		{author} {\bibfnamefont {W.}~\bibnamefont {Wegscheider}},\ and\ \bibinfo
		{author} {\bibfnamefont {L.~M.~K.}\ \bibnamefont {Vandersypen}},\ }\bibfield
	{title} {\bibinfo {title} {Coherent shuttle of electron-spin states},\ }\href
	{https://doi.org/10.1038/s41534-017-0024-4} {\bibfield  {journal} {\bibinfo
			{journal} {npj Quantum Information}\ }\textbf {\bibinfo {volume} {3}},\
		\bibinfo {pages} {22} (\bibinfo {year} {2017})}\BibitemShut {NoStop}%
	\bibitem [{\citenamefont {Flentje}\ \emph {et~al.}(2017)\citenamefont
		{Flentje}, \citenamefont {Mortemousque}, \citenamefont {Thalineau},
		\citenamefont {Ludwig}, \citenamefont {Wieck}, \citenamefont
		{B{\"{a}}uerle},\ and\ \citenamefont {Meunier}}]{Flentje2017}%
	\BibitemOpen
	\bibfield  {author} {\bibinfo {author} {\bibfnamefont {H.}~\bibnamefont
			{Flentje}}, \bibinfo {author} {\bibfnamefont {P.-A.}\ \bibnamefont
			{Mortemousque}}, \bibinfo {author} {\bibfnamefont {R.}~\bibnamefont
			{Thalineau}}, \bibinfo {author} {\bibfnamefont {A.}~\bibnamefont {Ludwig}},
		\bibinfo {author} {\bibfnamefont {A.~D.}\ \bibnamefont {Wieck}}, \bibinfo
		{author} {\bibfnamefont {C.}~\bibnamefont {B{\"{a}}uerle}},\ and\ \bibinfo
		{author} {\bibfnamefont {T.}~\bibnamefont {Meunier}},\ }\bibfield  {title}
	{\bibinfo {title} {{Coherent long-distance displacement of individual
				electron spins}},\ }\href {https://doi.org/10.1038/s41467-017-00534-3}
	{\bibfield  {journal} {\bibinfo  {journal} {Nat. Commun.}\ }\textbf {\bibinfo
			{volume} {8}},\ \bibinfo {pages} {501} (\bibinfo {year} {2017})}\BibitemShut
	{NoStop}%
	\bibitem [{\citenamefont {Leicht}\ \emph {et~al.}(2010)\citenamefont {Leicht},
		\citenamefont {Kaestner}, \citenamefont {Kashcheyevs}, \citenamefont
		{Mirovsky}, \citenamefont {Weimann}, \citenamefont {Pierz},\ and\
		\citenamefont {{H. W. }Schumacher}}]{Leicht2010}%
	\BibitemOpen
	\bibfield  {author} {\bibinfo {author} {\bibfnamefont {C.}~\bibnamefont
			{Leicht}}, \bibinfo {author} {\bibfnamefont {B.}~\bibnamefont {Kaestner}},
		\bibinfo {author} {\bibfnamefont {V.}~\bibnamefont {Kashcheyevs}}, \bibinfo
		{author} {\bibfnamefont {P.}~\bibnamefont {Mirovsky}}, \bibinfo {author}
		{\bibfnamefont {T.}~\bibnamefont {Weimann}}, \bibinfo {author} {\bibfnamefont
			{K.}~\bibnamefont {Pierz}},\ and\ \bibinfo {author} {\bibnamefont {{H. W.
				}Schumacher}},\ }\bibfield  {title} {\bibinfo {title} {Non-adiabatic pumping
			of single electrons affected by magnetic fields},\ }\href
	{https://doi.org/10.1016/j.physe.2009.11.109} {\bibfield  {journal} {\bibinfo
			{journal} {Physica E}\ }\textbf {\bibinfo {volume} {42}},\ \bibinfo {pages}
		{911 } (\bibinfo {year} {2010})},\ \bibinfo {note} {18th International
		Conference on Electron Properties of Two-Dimensional Systems}\BibitemShut
	{NoStop}%
	\bibitem [{\citenamefont {Amasha}\ \emph {et~al.}(2008)\citenamefont {Amasha},
		\citenamefont {MacLean}, \citenamefont {Radu}, \citenamefont {Zumb\"uhl},
		\citenamefont {Kastner}, \citenamefont {Hanson},\ and\ \citenamefont
		{Gossard}}]{Amasha2008}%
	\BibitemOpen
	\bibfield  {author} {\bibinfo {author} {\bibfnamefont {S.}~\bibnamefont
			{Amasha}}, \bibinfo {author} {\bibfnamefont {K.}~\bibnamefont {MacLean}},
		\bibinfo {author} {\bibfnamefont {I.~P.}\ \bibnamefont {Radu}}, \bibinfo
		{author} {\bibfnamefont {D.~M.}\ \bibnamefont {Zumb\"uhl}}, \bibinfo {author}
		{\bibfnamefont {M.~A.}\ \bibnamefont {Kastner}}, \bibinfo {author}
		{\bibfnamefont {M.~P.}\ \bibnamefont {Hanson}},\ and\ \bibinfo {author}
		{\bibfnamefont {A.~C.}\ \bibnamefont {Gossard}},\ }\bibfield  {title}
	{\bibinfo {title} {Spin-dependent tunneling of single electrons into an empty
			quantum dot},\ }\href {https://doi.org/10.1103/PhysRevB.78.041306} {\bibfield
		{journal} {\bibinfo  {journal} {Phys. Rev. B}\ }\textbf {\bibinfo {volume}
			{78}},\ \bibinfo {pages} {041306} (\bibinfo {year} {2008})}\BibitemShut
	{NoStop}%
	\bibitem [{\citenamefont {Stein}\ \emph {et~al.}(2015)\citenamefont {Stein},
		\citenamefont {Drung}, \citenamefont {Fricke}, \citenamefont {Scherer},
		\citenamefont {Hohls}, \citenamefont {Leicht}, \citenamefont {G\"otz},
		\citenamefont {Krause}, \citenamefont {Behr}, \citenamefont {Pesel},
		\citenamefont {Pierz}, \citenamefont {Siegner}, \citenamefont {Ahlers},\ and\
		\citenamefont {Schumacher}}]{Stein2015}%
	\BibitemOpen
	\bibfield  {author} {\bibinfo {author} {\bibfnamefont {F.}~\bibnamefont
			{Stein}}, \bibinfo {author} {\bibfnamefont {D.}~\bibnamefont {Drung}},
		\bibinfo {author} {\bibfnamefont {L.}~\bibnamefont {Fricke}}, \bibinfo
		{author} {\bibfnamefont {H.}~\bibnamefont {Scherer}}, \bibinfo {author}
		{\bibfnamefont {F.}~\bibnamefont {Hohls}}, \bibinfo {author} {\bibfnamefont
			{C.}~\bibnamefont {Leicht}}, \bibinfo {author} {\bibfnamefont
			{M.}~\bibnamefont {G\"otz}}, \bibinfo {author} {\bibfnamefont
			{C.}~\bibnamefont {Krause}}, \bibinfo {author} {\bibfnamefont
			{R.}~\bibnamefont {Behr}}, \bibinfo {author} {\bibfnamefont {E.}~\bibnamefont
			{Pesel}}, \bibinfo {author} {\bibfnamefont {K.}~\bibnamefont {Pierz}},
		\bibinfo {author} {\bibfnamefont {U.}~\bibnamefont {Siegner}}, \bibinfo
		{author} {\bibfnamefont {F.~J.}\ \bibnamefont {Ahlers}},\ and\ \bibinfo
		{author} {\bibfnamefont {H.~W.}\ \bibnamefont {Schumacher}},\ }\bibfield
	{title} {\bibinfo {title} {Validation of a quantized-current source with 0.2
			ppm uncertainty},\ }\href {https://doi.org/10.1063/1.4930142} {\bibfield
		{journal} {\bibinfo  {journal} {Appl. Phys. Lett.}\ }\textbf {\bibinfo
			{volume} {107}},\ \bibinfo {eid} {103501} (\bibinfo {year}
		{2015})}\BibitemShut {NoStop}%
	\bibitem [{\citenamefont {Pioro-Ladri\`ere}\ \emph {et~al.}(2005)\citenamefont
		{Pioro-Ladri\`ere}, \citenamefont {Davies}, \citenamefont {Long},
		\citenamefont {Sachrajda}, \citenamefont {Gaudreau}, \citenamefont
		{Zawadzki}, \citenamefont {Lapointe}, \citenamefont {Gupta}, \citenamefont
		{Wasilewski},\ and\ \citenamefont {Studenikin}}]{Pioro-Ladriere2005}%
	\BibitemOpen
	\bibfield  {author} {\bibinfo {author} {\bibfnamefont {M.}~\bibnamefont
			{Pioro-Ladri\`ere}}, \bibinfo {author} {\bibfnamefont {J.~H.}\ \bibnamefont
			{Davies}}, \bibinfo {author} {\bibfnamefont {A.~R.}\ \bibnamefont {Long}},
		\bibinfo {author} {\bibfnamefont {A.~S.}\ \bibnamefont {Sachrajda}}, \bibinfo
		{author} {\bibfnamefont {L.}~\bibnamefont {Gaudreau}}, \bibinfo {author}
		{\bibfnamefont {P.}~\bibnamefont {Zawadzki}}, \bibinfo {author}
		{\bibfnamefont {J.}~\bibnamefont {Lapointe}}, \bibinfo {author}
		{\bibfnamefont {J.}~\bibnamefont {Gupta}}, \bibinfo {author} {\bibfnamefont
			{Z.}~\bibnamefont {Wasilewski}},\ and\ \bibinfo {author} {\bibfnamefont
			{S.}~\bibnamefont {Studenikin}},\ }\bibfield  {title} {\bibinfo {title}
		{Origin of switching noise in {GaAs}/{AlxGa1-xAs} lateral gated devices},\
	}\href {https://doi.org/10.1103/PhysRevB.72.115331} {\bibfield  {journal}
		{\bibinfo  {journal} {Phys. Rev. B}\ }\textbf {\bibinfo {volume} {72}},\
		\bibinfo {pages} {115331} (\bibinfo {year} {2005})}\BibitemShut {NoStop}%
	\bibitem [{\citenamefont {Kashcheyevs}\ and\ \citenamefont
		{Kaestner}(2010)}]{Kashcheyevs2010}%
	\BibitemOpen
	\bibfield  {author} {\bibinfo {author} {\bibfnamefont {V.}~\bibnamefont
			{Kashcheyevs}}\ and\ \bibinfo {author} {\bibfnamefont {B.}~\bibnamefont
			{Kaestner}},\ }\bibfield  {title} {\bibinfo {title} {Universal decay cascade
			model for dynamic quantum dot initialization},\ }\href
	{https://doi.org/10.1103/PhysRevLett.104.186805} {\bibfield  {journal}
		{\bibinfo  {journal} {Phys. Rev. Lett.}\ }\textbf {\bibinfo {volume} {104}},\
		\bibinfo {pages} {186805} (\bibinfo {year} {2010})}\BibitemShut {NoStop}%
	\bibitem [{\citenamefont {Fricke}\ \emph {et~al.}(2013)\citenamefont {Fricke},
		\citenamefont {Wulf}, \citenamefont {Kaestner}, \citenamefont {Kashcheyevs},
		\citenamefont {Timoshenko}, \citenamefont {Nazarov}, \citenamefont {Hohls},
		\citenamefont {Mirovsky}, \citenamefont {Mackrodt}, \citenamefont {Dolata},
		\citenamefont {Weimann}, \citenamefont {Pierz},\ and\ \citenamefont
		{Schumacher}}]{Fricke2013}%
	\BibitemOpen
	\bibfield  {author} {\bibinfo {author} {\bibfnamefont {L.}~\bibnamefont
			{Fricke}}, \bibinfo {author} {\bibfnamefont {M.}~\bibnamefont {Wulf}},
		\bibinfo {author} {\bibfnamefont {B.}~\bibnamefont {Kaestner}}, \bibinfo
		{author} {\bibfnamefont {V.}~\bibnamefont {Kashcheyevs}}, \bibinfo {author}
		{\bibfnamefont {J.}~\bibnamefont {Timoshenko}}, \bibinfo {author}
		{\bibfnamefont {P.}~\bibnamefont {Nazarov}}, \bibinfo {author} {\bibfnamefont
			{F.}~\bibnamefont {Hohls}}, \bibinfo {author} {\bibfnamefont
			{P.}~\bibnamefont {Mirovsky}}, \bibinfo {author} {\bibfnamefont
			{B.}~\bibnamefont {Mackrodt}}, \bibinfo {author} {\bibfnamefont
			{R.}~\bibnamefont {Dolata}}, \bibinfo {author} {\bibfnamefont
			{T.}~\bibnamefont {Weimann}}, \bibinfo {author} {\bibfnamefont
			{K.}~\bibnamefont {Pierz}},\ and\ \bibinfo {author} {\bibfnamefont {H.~W.}\
			\bibnamefont {Schumacher}},\ }\bibfield  {title} {\bibinfo {title} {Counting
			statistics for electron capture in a dynamic quantum dot},\ }\href
	{https://doi.org/10.1103/PhysRevLett.110.126803} {\bibfield  {journal}
		{\bibinfo  {journal} {Phys. Rev. Lett.}\ }\textbf {\bibinfo {volume} {110}},\
		\bibinfo {pages} {126803} (\bibinfo {year} {2013})}\BibitemShut {NoStop}%
	\bibitem [{\citenamefont {Yamahata}\ \emph {et~al.}(2014)\citenamefont
		{Yamahata}, \citenamefont {Nishiguchi},\ and\ \citenamefont
		{Fujiwara}}]{Yamahata2014a}%
	\BibitemOpen
	\bibfield  {author} {\bibinfo {author} {\bibfnamefont {G.}~\bibnamefont
			{Yamahata}}, \bibinfo {author} {\bibfnamefont {K.}~\bibnamefont
			{Nishiguchi}},\ and\ \bibinfo {author} {\bibfnamefont {A.}~\bibnamefont
			{Fujiwara}},\ }\bibfield  {title} {\bibinfo {title} {Accuracy evaluation and
			mechanism crossover of single-electron transfer in si tunable-barrier
			turnstiles},\ }\href {https://doi.org/10.1103/PhysRevB.89.165302} {\bibfield
		{journal} {\bibinfo  {journal} {Phys. Rev. B}\ }\textbf {\bibinfo {volume}
			{89}},\ \bibinfo {pages} {165302} (\bibinfo {year} {2014})}\BibitemShut
	{NoStop}%
	\bibitem [{\citenamefont {Rossi}\ \emph {et~al.}(2018)\citenamefont {Rossi},
		\citenamefont {Klochan}, \citenamefont {Timoshenko}, \citenamefont {Hudson},
		\citenamefont {M{\"{o}}tt{\"{o}}nen}, \citenamefont {Rogge}, \citenamefont
		{Dzurak}, \citenamefont {Kashcheyevs},\ and\ \citenamefont
		{Tettamanzi}}]{Rossi2018}%
	\BibitemOpen
	\bibfield  {author} {\bibinfo {author} {\bibfnamefont {A.}~\bibnamefont
			{Rossi}}, \bibinfo {author} {\bibfnamefont {J.}~\bibnamefont {Klochan}},
		\bibinfo {author} {\bibfnamefont {J.}~\bibnamefont {Timoshenko}}, \bibinfo
		{author} {\bibfnamefont {F.~E.}\ \bibnamefont {Hudson}}, \bibinfo {author}
		{\bibfnamefont {M.}~\bibnamefont {M{\"{o}}tt{\"{o}}nen}}, \bibinfo {author}
		{\bibfnamefont {S.}~\bibnamefont {Rogge}}, \bibinfo {author} {\bibfnamefont
			{A.~S.}\ \bibnamefont {Dzurak}}, \bibinfo {author} {\bibfnamefont
			{V.}~\bibnamefont {Kashcheyevs}},\ and\ \bibinfo {author} {\bibfnamefont
			{G.~C.}\ \bibnamefont {Tettamanzi}},\ }\bibfield  {title} {\bibinfo {title}
		{{Gigahertz Single-Electron Pumping Mediated by Parasitic States}},\ }\href
	{https://doi.org/10.1021/acs.nanolett.8b00874} {\bibfield  {journal}
		{\bibinfo  {journal} {Nano Lett.}\ }\textbf {\bibinfo {volume} {18}},\
		\bibinfo {pages} {4141} (\bibinfo {year} {2018})}\BibitemShut {NoStop}%
	\bibitem [{\citenamefont {MacLean}\ \emph {et~al.}(2007)\citenamefont
		{MacLean}, \citenamefont {Amasha}, \citenamefont {Radu}, \citenamefont
		{Zumb{\"{u}}hl}, \citenamefont {Kastner}, \citenamefont {Hanson},\ and\
		\citenamefont {Gossard}}]{MacLean2007}%
	\BibitemOpen
	\bibfield  {author} {\bibinfo {author} {\bibfnamefont {K.}~\bibnamefont
			{MacLean}}, \bibinfo {author} {\bibfnamefont {S.}~\bibnamefont {Amasha}},
		\bibinfo {author} {\bibfnamefont {I.~P.}\ \bibnamefont {Radu}}, \bibinfo
		{author} {\bibfnamefont {D.~M.}\ \bibnamefont {Zumb{\"{u}}hl}}, \bibinfo
		{author} {\bibfnamefont {M.~A.}\ \bibnamefont {Kastner}}, \bibinfo {author}
		{\bibfnamefont {M.~P.}\ \bibnamefont {Hanson}},\ and\ \bibinfo {author}
		{\bibfnamefont {A.~C.}\ \bibnamefont {Gossard}},\ }\bibfield  {title}
	{\bibinfo {title} {{Energy-dependent tunneling in a quantum dot}},\ }\href
	{https://doi.org/10.1103/PhysRevLett.98.036802} {\bibfield  {journal}
		{\bibinfo  {journal} {Phys. Rev. Lett.}\ }\textbf {\bibinfo {volume} {98}},\
		\bibinfo {pages} {036802} (\bibinfo {year} {2007})}\BibitemShut {NoStop}%
	\bibitem [{sup()}]{supp}%
	\BibitemOpen
	\href@noop {} {}\bibinfo {note} {See Supplemental Material for details on
		experimental setup, fitting and simulation procedures, and additional
		measurements.}\BibitemShut {Stop}%
	\bibitem [{\citenamefont {Stano}\ and\ \citenamefont
		{Fabian}(2006)}]{Stano2006}%
	\BibitemOpen
	\bibfield  {author} {\bibinfo {author} {\bibfnamefont {P.}~\bibnamefont
			{Stano}}\ and\ \bibinfo {author} {\bibfnamefont {J.}~\bibnamefont {Fabian}},\
	}\bibfield  {title} {\bibinfo {title} {Orbital and spin relaxation in single
			and coupled quantum dots},\ }\href
	{https://doi.org/10.1103/PhysRevB.74.045320} {\bibfield  {journal} {\bibinfo
			{journal} {Phys. Rev. B}\ }\textbf {\bibinfo {volume} {74}},\ \bibinfo
		{pages} {045320} (\bibinfo {year} {2006})}\BibitemShut {NoStop}%
	\bibitem [{\citenamefont {Hanson}\ \emph {et~al.}(2007)\citenamefont {Hanson},
		\citenamefont {Kouwenhoven}, \citenamefont {Petta}, \citenamefont {Tarucha},\
		and\ \citenamefont {Vandersypen}}]{Hanson2007}%
	\BibitemOpen
	\bibfield  {author} {\bibinfo {author} {\bibfnamefont {R.}~\bibnamefont
			{Hanson}}, \bibinfo {author} {\bibfnamefont {L.~P.}\ \bibnamefont
			{Kouwenhoven}}, \bibinfo {author} {\bibfnamefont {J.~R.}\ \bibnamefont
			{Petta}}, \bibinfo {author} {\bibfnamefont {S.}~\bibnamefont {Tarucha}},\
		and\ \bibinfo {author} {\bibfnamefont {L.~M.~K.}\ \bibnamefont
			{Vandersypen}},\ }\bibfield  {title} {\bibinfo {title} {Spins in few-electron
			quantum dots},\ }\href {https://doi.org/10.1103/RevModPhys.79.1217}
	{\bibfield  {journal} {\bibinfo  {journal} {Rev. Mod. Phys.}\ }\textbf
		{\bibinfo {volume} {79}},\ \bibinfo {pages} {1217} (\bibinfo {year}
		{2007})}\BibitemShut {NoStop}%
	\bibitem [{\citenamefont {Fujisawa}\ \emph {et~al.}(2002)\citenamefont
		{Fujisawa}, \citenamefont {Austing}, \citenamefont {Tokura}, \citenamefont
		{Hirayama},\ and\ \citenamefont {Tarucha}}]{Fujisawa2002}%
	\BibitemOpen
	\bibfield  {author} {\bibinfo {author} {\bibfnamefont {T.}~\bibnamefont
			{Fujisawa}}, \bibinfo {author} {\bibfnamefont {D.~G.}\ \bibnamefont
			{Austing}}, \bibinfo {author} {\bibfnamefont {Y.}~\bibnamefont {Tokura}},
		\bibinfo {author} {\bibfnamefont {Y.}~\bibnamefont {Hirayama}},\ and\
		\bibinfo {author} {\bibfnamefont {S.}~\bibnamefont {Tarucha}},\ }\bibfield
	{title} {\bibinfo {title} {Allowed and forbidden transitions in artificial
			hydrogen and helium atoms},\ }\href {https://doi.org/10.1038/nature00976}
	{\bibfield  {journal} {\bibinfo  {journal} {Nature}\ }\textbf {\bibinfo
			{volume} {419}},\ \bibinfo {pages} {278} (\bibinfo {year}
		{2002})}\BibitemShut {NoStop}%
	\bibitem [{\citenamefont {Sasaki}\ \emph {et~al.}(2005)\citenamefont {Sasaki},
		\citenamefont {Fujisawa}, \citenamefont {Hayashi},\ and\ \citenamefont
		{Hirayama}}]{Sasaki2005}%
	\BibitemOpen
	\bibfield  {author} {\bibinfo {author} {\bibfnamefont {S.}~\bibnamefont
			{Sasaki}}, \bibinfo {author} {\bibfnamefont {T.}~\bibnamefont {Fujisawa}},
		\bibinfo {author} {\bibfnamefont {T.}~\bibnamefont {Hayashi}},\ and\ \bibinfo
		{author} {\bibfnamefont {Y.}~\bibnamefont {Hirayama}},\ }\bibfield  {title}
	{\bibinfo {title} {Electrical pump-and-probe study of spin singlet-triplet
			relaxation in a quantum dot},\ }\href
	{https://doi.org/10.1103/PhysRevLett.95.056803} {\bibfield  {journal}
		{\bibinfo  {journal} {Phys. Rev. Lett.}\ }\textbf {\bibinfo {volume} {95}},\
		\bibinfo {pages} {056803} (\bibinfo {year} {2005})}\BibitemShut {NoStop}%
	\bibitem [{\citenamefont {Hanson}\ \emph {et~al.}(2005)\citenamefont {Hanson},
		\citenamefont {van Beveren}, \citenamefont {Vink}, \citenamefont {Elzerman},
		\citenamefont {Naber}, \citenamefont {Koppens}, \citenamefont {Kouwenhoven},\
		and\ \citenamefont {Vandersypen}}]{Hanson2005}%
	\BibitemOpen
	\bibfield  {author} {\bibinfo {author} {\bibfnamefont {R.}~\bibnamefont
			{Hanson}}, \bibinfo {author} {\bibfnamefont {L.~H.}\ \bibnamefont {Willems van
				Beveren}}, \bibinfo {author} {\bibfnamefont {I.~T.}\ \bibnamefont {Vink}},
		\bibinfo {author} {\bibfnamefont {J.~M.}\ \bibnamefont {Elzerman}}, \bibinfo
		{author} {\bibfnamefont {W.~J.~M.}\ \bibnamefont {Naber}}, \bibinfo {author}
		{\bibfnamefont {F.~H.~L.}\ \bibnamefont {Koppens}}, \bibinfo {author}
		{\bibfnamefont {L.~P.}\ \bibnamefont {Kouwenhoven}},\ and\ \bibinfo {author}
		{\bibfnamefont {L.~M.~K.}\ \bibnamefont {Vandersypen}},\ }\bibfield  {title}
	{\bibinfo {title} {Single-shot readout of electron spin states in a quantum
			dot using spin-dependent tunnel rates},\ }\href
	{https://doi.org/10.1103/PhysRevLett.94.196802} {\bibfield  {journal}
		{\bibinfo  {journal} {Phys. Rev. Lett.}\ }\textbf {\bibinfo {volume} {94}},\
		\bibinfo {pages} {196802} (\bibinfo {year} {2005})}\BibitemShut {NoStop}%
	\bibitem [{\citenamefont {Fletcher}\ \emph {et~al.}(2013)\citenamefont
		{Fletcher}, \citenamefont {See}, \citenamefont {Howe}, \citenamefont
		{Pepper}, \citenamefont {Giblin}, \citenamefont {Griffiths}, \citenamefont
		{Jones}, \citenamefont {Farrer}, \citenamefont {Ritchie}, \citenamefont
		{Janssen},\ and\ \citenamefont {Kataoka}}]{Fletcher2013}%
	\BibitemOpen
	\bibfield  {author} {\bibinfo {author} {\bibfnamefont {J.~D.}\ \bibnamefont
			{Fletcher}}, \bibinfo {author} {\bibfnamefont {P.}~\bibnamefont {See}},
		\bibinfo {author} {\bibfnamefont {H.}~\bibnamefont {Howe}}, \bibinfo {author}
		{\bibfnamefont {M.}~\bibnamefont {Pepper}}, \bibinfo {author} {\bibfnamefont
			{S.~P.}\ \bibnamefont {Giblin}}, \bibinfo {author} {\bibfnamefont {J.~P.}\
			\bibnamefont {Griffiths}}, \bibinfo {author} {\bibfnamefont {G.~A.~C.}\
			\bibnamefont {Jones}}, \bibinfo {author} {\bibfnamefont {I.}~\bibnamefont
			{Farrer}}, \bibinfo {author} {\bibfnamefont {D.~A.}\ \bibnamefont {Ritchie}},
		\bibinfo {author} {\bibfnamefont {T.~J. B.~M.}\ \bibnamefont {Janssen}},\
		and\ \bibinfo {author} {\bibfnamefont {M.}~\bibnamefont {Kataoka}},\
	}\bibfield  {title} {\bibinfo {title} {Clock-controlled emission of
			single-electron wave packets in a solid-state circuit},\ }\href
	{https://doi.org/10.1103/PhysRevLett.111.216807} {\bibfield  {journal}
		{\bibinfo  {journal} {Phys. Rev. Lett.}\ }\textbf {\bibinfo {volume} {111}},\
		\bibinfo {pages} {216807} (\bibinfo {year} {2013})}\BibitemShut {NoStop}%
\end{thebibliography}

%

\newcommand{\figOne}{\ref{fig:device}}
\newcommand{\figTwo}{\ref{fig:oneemap}}
\newcommand{\figThree}{\ref{fig:fits}}
\newcommand{\figFour}{\ref{fig:twoemaps}}

\newcommand{\eqOne}{\ref{eq:I}}
\newcommand{\eqTwo}{\ref{eq:gammain}}
\newcommand{\eqThree}{\ref{eq:gamma}}

\newcommand{\refStein}{\onlinecite{Stein2017}}
\newcommand{\refRossi}{\onlinecite{Rossi2018}}
\newcommand{\refKaestner}{\onlinecite{Kaestner2015}}

\setcounter{equation}{0}
\setcounter{figure}{0}
\renewcommand{\theequation}{S\arabic{equation}}
\renewcommand{\thefigure}{S\arabic{figure}}
\renewcommand{\thetable}{S\arabic{table}}


\vspace{2cm}
\begin{center}
\textbf{\Large Supplemental Material}
\end{center}

\section{Experimental details}

\begin{figure}[b]
	\includegraphics[width=0.75\columnwidth]{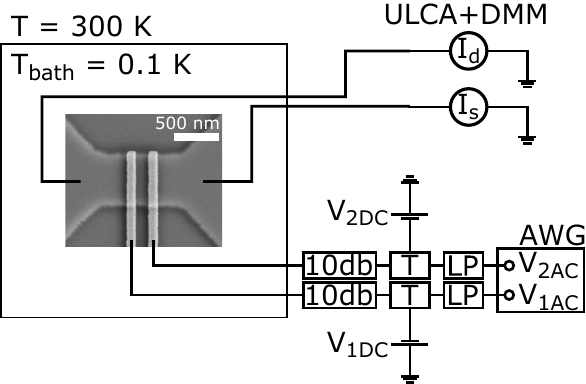} 
	\caption{A simplified scheme of the measurement setup.}
	\label{fig:setup}
\end{figure}
The measurements where performed in a top-loading dilution refrigerator. The device is immersed in the $^3$He/$^4$He mixture. The gates are connected to semi-rigid coaxial lines within the top loading probe (about 2.8 meter length) to allow application of high bandwidth signals. The time dependent wave form is generated by a 10 GS/s arbitrary waveform generator (AWG, Tektronix 7000 series). The signal is filtered by a low pass filter (Mini-Circuits SBLT-1870+) to remove digitizing steps and overshoots. For fine control of the DC offset a separate custom made digital-analog converter (DAC) source is used. The DAC and AWG signals are combined by a bias-T (Mini-Circuits ZFBT-6G+) which is connected by flexible coaxial cables to a 10 dB attenuator screwed directly to the RF port of the top loading port (see Fig.~\ref{fig:setup} for simplified setup). The measured attenuation is 0.55. This attenuation and the AC coupling of the AWG was taken into account to calculate all voltage values shown.

Due to heating considerations no matched load resistor is connected to the low temperature end of the coaxial line. Thus the coaxial line is terminated only by a small parasitic capacitance of the device. Due to this and due to the non-perfect voltage standing wave ratio of the attenuator, bias-T, and AWG, small echoes of large sudden voltage changes reach the device after 28 ns and (a weaker one) after 48 ns. These echoes can disturb the measurement when occurring in a critical phase of the signal cycle. For the measurements shown in the main text this was prevented by introducing a 50 ns waiting time between the ejection phase of the cycle, phase (iii), and the loading phase (i) in the applied 
time-dependent entry gate voltage waveform displayed in Fig.~\figOne (b). The voltage level during this waiting time was chosen such that the ground state is still above the Fermi level of the source and no loading occurs. The other durations of the signal cycle are 3.3 ns for phase (ii), 1.7 ns for phase (iii) and a varying time $T_{\text{load}}$ as indicated in the main text for the loading phase (i). The voltage change from phase (iii) to waiting level is +522.5 mV, the voltage change from waiting level to loading only +27.5 mV, i.e. 19 times smaller. As a result the echoes of the big step are absorbed in the waiting time without influence on the measurement, while the echoes of the second step into the loading phase are sufficiently small not to disturb the measurement.

The generated current is measured both in the source and the drain lead by current to voltage conversion by transimpedance amplifiers (type ULCA, see e.g.\ Ref.~\refStein\,  and references therein for technical details) and voltage measurements using two digital multimeters (DMMs, Agilent 3458A).

\begin{figure}[b]
	\includegraphics[width=0.85\columnwidth]{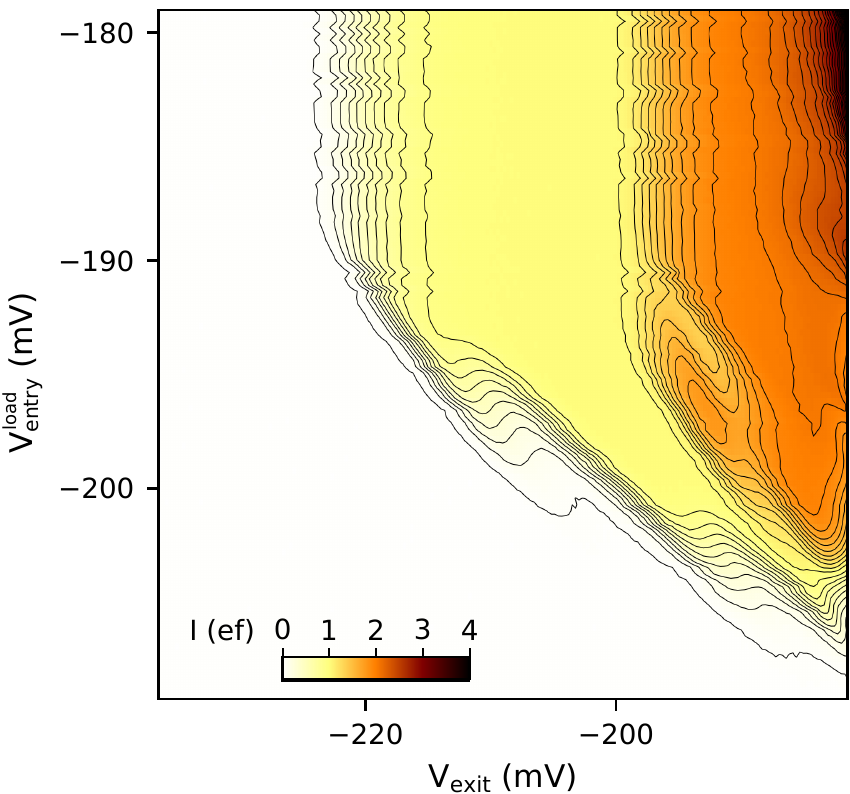} 
	\caption{Average captured charge for a second device, measured without waiting plateau between the ejection phase (iii) and the loading phase (i). $T_{\text{load}}=50$~ns.}
	\label{fig:sampl2}
\end{figure}

Another device was measured preceding the measurements analyzed in the paper. These preceding measurements were performed without additional waiting time. A plot equivalent to the measurement shown in Fig.~\figTwo\ of the main text is shown in Fig.~\ref{fig:sampl2}. All key features discussed in the text are reproduced in this measurement, however, the dataset shows indications of voltage shifts due to a two-level fluctuator. The selection of the device and the dataset for the analysis was based on having used the additional wait time to avoid echo influences and on better device stability.

\section{Fitting procedure for\\ the gate-voltage dependence of\\ the tunneling rates}
\begin{enumerate}
	\item Approximate initial value for the constant energy slope $s_\mu$ is chosen.
	\item A set of crossections
	$\VLE +s_{\mu}\, \Vexit = \Delta V$ is selected for a range of $\Delta V$ that covers the SE capture region in the 2D plot of $\langle {n} \rangle(\VLE, \Vexit)$.
	\item For a fixed $s_{\mu}$, and for each $\Delta V$, one-dimensional least-squares fits to Eq.~(\eqOne) of the main text,
	\begin{multline}\label{eq:actualfit}
	\langle {n} \rangle(\Delta V\! -\! s_{\mu}\Vexit, \Vexit) =  \\
	= \exp(-e^{\!-\!\alpha \Vexit+\delta_{\text{back}}}) 
	\left[
	1-\exp \left( -e^{-\alpha \Vexit+\delta_{\text{load}}}\right)\right ] \, ,
	\end{multline}
	are performed. The adjustable values of $\alpha$ and $\delta_{\text{back}}$ are the same for all crosssections, while $\delta_{\text{load}}(\Delta V)$ is allowed to vary from slice to slice.
	\item The value of $s_{\mu}$ is optimised by repeating step 2 and 3 above, until the sum of the residuals
	for all one-dimensional fits is minimised.
\end{enumerate}
The values of $s_{\mu} =0.603$ and $\alpha = 0.169 \, \text{mV}^{-1}$ have been determined by the above algorithm from the maps $\langle {n} \rangle(\VLE, \Vexit)$ measured at $\tload = 50 \,  \text{ns}$ and $\mathrm{T}_{\text{bath}} = 0.1 \, \text{K}$ (and shown in Fig.~\figTwo). These values of $s_{\mu}$ and $\alpha$ are used for all subsequent rate measurements and captured electron number simulations.

The values of $W_{\text{in}}(\Delta V-s_{\mu} \Vexit^{0})=\tload^{-1} \exp ( \delta_{\text{load}}-\alpha \Vexit^{0})$ are computed from the fitted values $\delta_{\text{load}} (\Delta V)$, thus yielding the function $W_{\text{in}}$  defined in Eq.~(\eqTwo) and plotted in Fig.~\figThree.

\section{Simulation details and\\ parameter values}
The basis for modelling the loading dynamics is the gate-voltage dependence of intunneling rates,
parametrized as 
\begin{align} \label{eqref:GammaIndividual}
\Gamma_{n,i}\left( \VLE, \Vexit \right) & =
\gamma_{n,i}\left( \VLE, \Vexit \right)  \\
& \times \, f\left(\VLE \!+\!s_{\mu} \Delta \Vexit-\Ventry^{n,i}\right) \, , 
\nonumber \\
\gamma_{n,i}^{} \left( \VLE, \Vexit \right) &= \gamma_{n,i}^0 \, e^{k_\gamma \VLE-\alpha \Delta \Vexit } \, .
\end{align}
For $n=1$, the sum $\Gamma_{\text{in}}= \sum_{i=0}^2 \Gamma_{1,i}$ is the total single-electron tunneling
rate $\Gamma_{\text{in}}$ discussed in the main text.
The lever-arm parameters $\alpha$ and $s_{\mu}$ are determined by the fitting procedure described earlier,
the thresholds $\Ventry^{1,i}$ and the rate ratios $\gamma_{1,i}/\gamma_{1,0}$ for $i=1$ and $2$ indicated in Table~\ref{tab:tab1} are obtained by fitting Eq.~(\eqThree) of the main text to the experimentally determined $W_{\text{in}}$. 
\begin{table}[tb]
	\caption{Values of discrete state parameters used in the simulations.\label{tab:tab1}}
	\begin{center}
		\begin{ruledtabular}
			\begin{tabular}{c|c|c|c|d|d} 
				$(n, i)$ &  type & relaxes to & $g_{n,i}$  &
				\multicolumn{1}{c|}{ $\Ventry^{n,i}-\Ventry^{1,0}$, \text{mV}}
				& \multicolumn{1}{c}{ $\ln (\gamma_{n,i}/\gamma_{1,0})$ }
				\\  \hline
				(1, 0) &  1e  & --- & 2  &  0 
				&  0 \\ 
				(1, 1)  &  1e   & (1,0) &  2  & 7.57 
				&  5.05\footnotemark[1] \\ 
				(1, 2)  &  1e   & (1,0) & 2  &  11.6 
				&  7.53\footnotemark[1] \\ 
				(1, 3)  &  1e   & (1,0) &  2  &  18.1 
				&  15.0\footnotemark[2] \\ \hline
				(2, 0)  &  S   & --- &  1  & 14.4 
				&  5.96 
				\\ 
				(2, 1)  &  T   & --- &  3  & 17.0  
				&  8.64 
				\\ 
				(2,  2)  &  S   & (2,0) &  1  &  24.7 
				&  15.0\footnotemark[2]  
			\end{tabular}
		\end{ruledtabular}
		\footnotetext{Values fixed by one-dimensional fit to Eq.~(\eqThree) in Fig.~\figThree(c).}
		\footnotetext{A sufficiently large value to guarantee loading once this highest modelled excited state is energetically available.}
	\end{center}
\end{table}
This fit also yields
$k_B \mathrm{T}_{\text{eff}}/k_\mu =0.38\,\text{mV}$ which we use subsequently.
More generally, $\Gamma_{n,i}$ would also depend exponentially on $\VLE$ at fixed $\Vexit$ with a scaling factor $e^{k_\gamma \VLE}$.  For the device described in the main text, we find the coupling parameter  $k_\gamma\approx0$ within the uncertainties of the fit, indicating that under variation of $\Ventry$, the energetic variation of the QD levels is $\mu_{n,i}$ is comparable to the variation of the effective barrier height.
Parameters of the first two excited single-electron states imply that an e-fold increase in $\gamma$ corresponds to $\Delta \mu/k_{\mu}\approx 1.5$ mV change in energy; $k_B \mathrm{T}_{\text{eff}} < \Delta \mu$ is consistent with direct tunnelling model and neglect of thermal hopping.

In principle, the instantaneous backtunneling rates $\Gamma_{n,i}^{\text{back}}$ can be predicted using detailed balance condition and Eq.~\eqref{eqref:GammaIndividual}. The resulting gate-voltage dependence $\Gamma_{n,i}^{\text{back}}$ would be similar to  Eq.~\eqref{eqref:GammaIndividual}, but with $f$ replaced by $1-f$ and  $\gamma_{n,i}^{}$ replaced by 
$\gamma_{n,i}^{\text{back}}=(g_{n,i}^{})^{-1} \gamma_{n,i}^{}$ where $g_{n,i}^{}$ is the quantum degeneracy of level $(n,i)$~[\refRossi]. 
We take a more direct, simplified approach, by approximating the ratio of logarithms of state-specific survival (non-backtunneliling) probabilities,
$P_{n,i}^{\text{N}} \equiv 1- P_{n,i}^{\text{back}} = \exp \left (- \int \Gamma_{n,i}^{\text{back}} dt \right )$, with the ratio of the corresponding rates, $(\ln  P_{n,i}^{\text{N}}) / (\ln P_{1,0}^{\text{N}}) \approx \gamma_{n,i}^{\text{back}}/\gamma_{1,0}^{\text{back}}$. Such an approach is justified when timescale for  closing the barrier is much larger than the delay in the onset of backtunneling due to energy difference  (large plunger-to-barrier ratio [\refKaestner]); this is consistent with the observed small $k_{\gamma}$.

Below we list specific equations used to produce data in Fig.~\figFour (a)-(c) and (e).
Corresponding numerical values of the parameters are listed in Table~\ref{tab:tab1}.
\subsection{(a) Single electron, single state}
We solve the equation for the loading probability 
\begin{align}
\frac{d}{d t} P_1^{\text{L}}(t) & = \Gamma_{1,0}\left( \VLE, \Vexit \right)[ 1-P_1^{\text{L}}(t)] 
\end{align}
with the initial condition $P_1^{\text{L}}(0)=0$. The average captured charge is computed as 
\begin{align}
\langle {n} \rangle= P_1^{\text{L}}(T_{\text{load}}) P_{1}^{\text{N}} \, ,
\end{align}
where    
\begin{align}
P_{1}^{\text{N}} & = 1- P_{\text{back}} =\exp\left (-e^{-\alpha \Vexit+\delta_{\text{back}}} \right ) \, ,
\end{align}
and a constant value of $\delta_{\text{back}}$ is used, consistent with the position the fits of the experimental data.

\subsection{(b) Single electron, multiple states, full relaxation}
Same as (a) but with the ground state loading rate $\Gamma_{1,0}$ replaced by total rate $\Gamma_{\text{in}}= \sum_{i=0}^2 \Gamma_{1,i}$.

\subsection{(c) Single electron, multiple states, no relaxation \label{sec:1enorelax}}
Three equations for $i=0,1$ and $2$ are solved,
\begin{align}
\frac{d}{d t} P_{1,i}^{\text{L}}(t) & = \Gamma_{1,i}\left( \VLE, \Vexit \right)[ 1-\sum_{j=0}^2 P_{1,j}^{\text{L}}(t)] \, , 
\end{align}
with the initial condition $P_{1,i}^{\text{L}}(0)=0$.  
The average captured charge is
\begin{align}
\langle {n} \rangle= \sum_{i=0}^2 P_{1,i}^{\text{L}}(T_{\text{load}}) P_{1,i}^{\text{N}} \, ,
\end{align}
where the survival probabilities are
\begin{align}
\ln P_{1,i}^{\text{N}} & =  \frac{\gamma_{1,i}^{\text{back}}}{\gamma_{1,0}^{\text{back}}}  \ln  P_{1}^{\text{N}} = \frac{g_{1,0} \gamma_{1,i}}{g_{1,i}  \gamma_{1,0}}  \ln  P_{1}^{\text{N}} \, .
\end{align}

\subsection{(e) Single and two electrons\label{sec:full}}
Probabilities $P_1^{\text{L}}$, $P_S^{\text{L}}$ and $P_T^{\text{L}}$ for loading into three distinct final states, $(1,0)$, $(2,0)$, and $(2,1)$, respectively, are computed from the rate equations
\begin{multline}
\frac{d}{d t} P_{1}^{\text{L}}(t)  = \Gamma_{\text{in}}\left( \VLE, \Vexit \right) \\
\times \left [ 1-P_{1}^{\text{L}}(t)-P_{S}^{\text{L}}(t)-P_{T}^{\text{L}}(t) \right] \\
-\left (\Gamma_{\text{in}}^{\text{S}}  +\Gamma_{\text{in}}^{\text{T}} \right) \, P_{1}^{\text{L}}(t)  \, ,
\end{multline}
\vspace{-0.8cm}
\begin{align}
\frac{d}{d t} P_{2S}^{\text{L}}(t) & = \Gamma_{\text{in}}^{\text{S}} \left( \VLE, \Vexit \right) \, P_{1}^{\text{L}}(t) \, , \\
\frac{d}{d t} P_{2T}^{\text{L}}(t) & =  \Gamma_{\text{in}}^{\text{T}} \left( \VLE, \Vexit \right) \, P_{1}^{\text{L}}(t) \, ,
\end{align}
with $P_1^{\text{L}}(0)=P_{S}^{\text{L}}(0) =P_{T}^{\text{L}}(0)=0$, 
$\Gamma_{\text{in}}^{\text{S}} =\Gamma_{2,0}+ \Gamma_{2,2}$,
$\Gamma_{\text{in}}^{\text{T}} = \Gamma_{2,1}$, and
$\Gamma_{\text{in}}= \sum_{i=0}^3 \Gamma_{1,i}$ including a fourth single-electron state.

The difference in backtunneling rates between $S$-type and $T$-type final states is reflected by the corresponding survival probabilities:
\begin{align}
\ln P_{S}^{\text{N}} & =  \frac{\gamma_{2,0}^{\text{back}}}{\gamma_{1,0}^{\text{back}}}  \ln  P_{1}^{\text{N}} = \frac{g_{1,0} \gamma_{2,0}}{g_{2,0}  \gamma_{1,0}}  \ln  P_{1}^{\text{N}} \, , \\
\ln P_{T}^{\text{N}} & =  \frac{\gamma_{2,1}^{\text{back}}}{\gamma_{1,0}^{\text{back}}}  \ln  P_{1}^{\text{N}} = \frac{g_{1,0} \gamma_{2,1}}{g_{2,1}  \gamma_{1,0}} \ln P_{1}^{\text{N}}  \, .
\end{align}
The parameters of the two-electron states in Table~\ref{tab:tab1} are chosen to achieve qualitative agreement with experimental data in Fig.~\figFour(d).
Note that $\gamma_{2,1}^{\text{back}}/\gamma_{2,0}^{\text{back}} =  ( \gamma_{2,1}/g_{2,1})/( \gamma_{2,0}/g_{2,0})=4.9$  in accord with the estimate quoted in the main text.

The total capture probability is computed as 
\begin{multline}
\langle {n} \rangle= 2 \, P_{S}^{\text{L}}(\tload)\, P_{S}^{\text{N}} + 2 \, P_{T}^{\text{L}}(\tload) \, P_{T}^{\text{N}} \\ 
+ [ P_{S}^{\text{L}}(\tload) (1- P_{S}^{\text{N}})+ P_{T}^{\text{L}}(\tload) (1- P_{T}^{\text{N}})+P_1^{\text{L}}(\tload) ] \, P_{1}^{\text{N}} \, . \label{eq:twoelectroncaptur}
\end{multline}

The term in the square brackets in Eq.~\eqref{eq:twoelectroncaptur} takes into account the possibility of initialising the single-electron state via the decay of a two-electron state: loading (probabilities $P_{S}^{\text{L}}$ and $P_{T}^{\text{L}}$ for S-type and T-type, respectively) but not keeping (probability $(1- P_{S}^{\text{N}})$ and $(1- P_{T}^{\text{N}})$, respectively) the second electron is analternative initialization route, in addition to direct loading into the single electron state (probability  $P_1^{\text{L}}$).

\section{Temperature dependence of\\ the SE intunnelling rate}
\begin{figure}[b]
	\includegraphics[width=0.9\columnwidth]{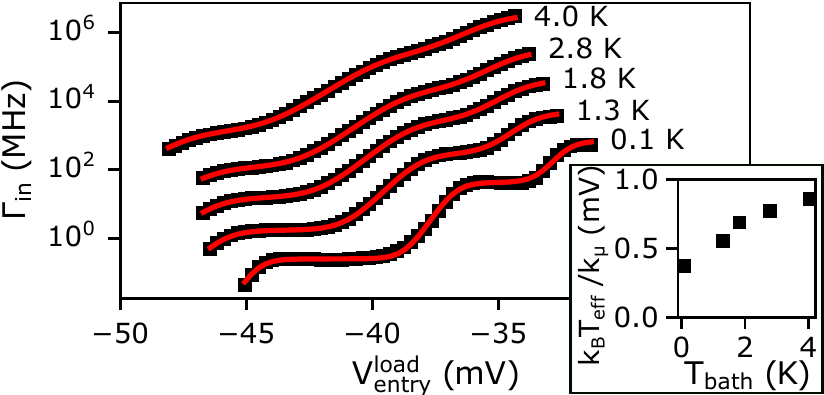}
	\caption{Temperature-dependent measurements of $\Gamma_{\mathrm{in}}$ (stacked with horizontal and vertical offset) show increased broadening of steps.
		Red lines are fits using Eq.~(\eqThree).
		Inset: Extracted effective temperature.}
	\label{fig:temperature}
\end{figure}

In order to confirm the assumption of Fermi-broadened transitions, we performed the experiment for various temperatures at fixed $\tload=9\,\mathrm{ns}$. 
Figure~\ref{fig:temperature} shows the result of the analysis, showing broadened steps with increasing temperature.
(Data for the lowest temperature $\text{T}_{\text{bath}}=0.1$~K shown here are not identical to the 9 ns data shown in Fig.~\figThree, the measurements were done at different times.)
The  effective temperature $\mathrm{T}_{\text{eff}}$ extracted by fitting to  Eq.~(\eqThree) is shown in the inset to Fig.~\ref{fig:temperature}. It shows a roughly linear dependence on the bath temperature.
If we assume $\mathrm{T}_{\text{eff}} = \mathrm{T}_{\text{bath}}$ at at the maximal $\mathrm{T}_{\text{bath}}=4.1\,\text{K}$ then the effective temperature corresponding to the measurements reported in the main text (performed at $\mathrm{T}_{\text{bath}}=100 \,\text{mK}$) is $\mathrm{T}_{\text{eff}}  \approx 1.4\,\text{K}$. Possible reasons for elevated effective temperature are excess dissipated radio frequency power and deviations from perfect flatness in the wave-form that keeps the gate-controlled energy levels $\mu_{n,i}-E_F$ constant during the loading phase.

\end{document}